%% file: main.tex
\documentclass[10pt,twocolumn,letterpaper]{article}

\usepackage[pagenumbers]{cvpr} 

\input{preamble}

\definecolor{cvprblue}{rgb}{0.21,0.49,0.74}
\usepackage[pagebackref,breaklinks,colorlinks,allcolors=cvprblue]{hyperref}

\def\confName{CVPR}
\def\confYear{2026}

\title{\projectname: Hyperspectral Video via 
\\Active Illumination and Coded-Exposure Pixels}

\author{
Dhruv Verma$^{1}$ \quad
Andrew Qiu$^{1}$ \quad
Roberto Rangel$^{2}$ \quad
Ayandev Barman$^{2}$ \quad
Hao Yang$^{2}$ \quad
Chenjia Hu$^{2}$ \\
Fengqi Zhang$^{2}$ \quad
Roman Genov$^{2}$ \quad
David B. Lindell$^{1}$ \quad
Kiriakos N. Kutulakos$^{1}$ \quad
Alex Mariakakis$^{1}$ \\
\\
$^{1}$Department of Computer Science, University of Toronto, Canada \\
$^{2}$Department of Electrical \& Computer Engineering, University of Toronto, Canada \\
{\tt\small dhruvverma@cs.toronto.edu}
}

\begin{document}

\input{sections/0_teaser_figure}
\maketitle

\input{sections/0_abstract}    
\input{sections/1_introduction}
\input{sections/2_related_work}
\input{sections/3_image_formation}
\input{sections/4_reconstruction}
\input{sections/5_evaluation}
\input{sections/6_conclusion}

\input{sections/7_long_figures}
{
    \small
    \bibliographystyle{ieeenat_fullname}
    \bibliography{ref}
}

\input{suppl}

\end{document}

%% file: preamble.tex
\usepackage{xspace}
\usepackage{multirow}
\usepackage{caption}
\usepackage{subcaption}
\usepackage{graphicx}
\usepackage{enumitem}
\usepackage{bbm}
\usepackage{textcomp}
\usepackage[dvipsnames]{xcolor}
\usepackage{tabularray}
\usepackage{booktabs}
\usepackage{gensymb}
\usepackage{graphicx}
\usepackage{longtable}
\usepackage{titletoc}
\usepackage{placeins}

\newcommand*{\projectname}{Lumosaic\xspace}
\newcommand{\redacted}[1]{\textbf{REDACTED}}

\newenvironment{s_itemize}{
\begin{itemize}[nosep,leftmargin=15pt]
}{\end{itemize}}
\newenvironment{s_enumerate}{
\begin{enumerate}[nosep,leftmargin=15pt]
}{\end{enumerate}}
\DeclareMathOperator*{\argmin}{\arg\min}

\emergencystretch=\maxdimen
\expandafter\def\expandafter\normalsize\expandafter{%
    \normalsize%
    \setlength\abovedisplayskip{3pt}%
    \setlength\belowdisplayskip{3pt}%
    \setlength\abovedisplayshortskip{3pt}%
    \setlength\belowdisplayshortskip{3pt}%
}



%% file: sections/0_teaser_figure.tex
\twocolumn[{%
  \renewcommand\twocolumn[1][]{#1}
  \maketitle
  \begin{center}
    \vspace{-6mm} 
    \includegraphics[width=\textwidth,height=1\textheight,keepaspectratio]{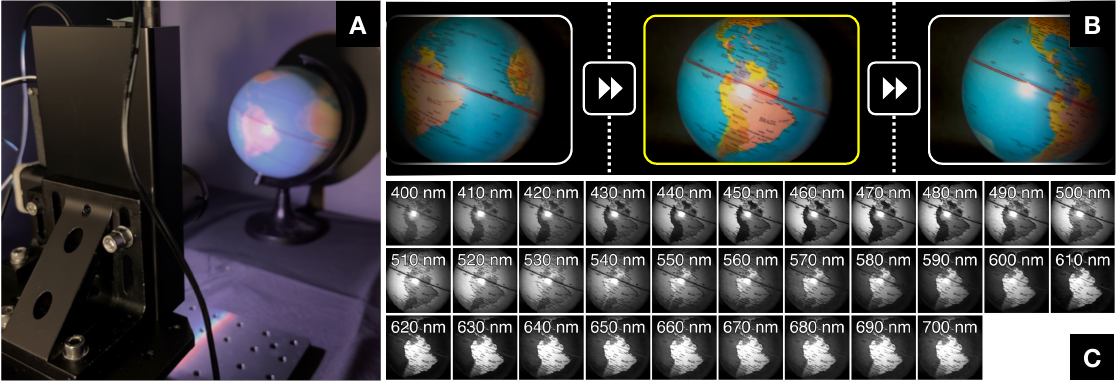}
    \captionof{figure}{
    \textbf{(A)} \projectname\ is a compact active hyperspectral video system that combines programmable narrowband illumination with pixel-wise coded exposure to jointly encode spatial, spectral, and temporal information within each video frame, enabling real-time capture at 30 fps.
    \textbf{(B)} Reconstructed frames of a rotating globe, rendered in sRGB, showing smooth temporal progression despite rapid motion.
    \textbf{(C)} Reconstructed spectral channels (400-700\,nm, 10\,nm intervals) from the highlighted frame.
    }
    \label{fig:teaser}
  \end{center}
}]

%% file: sections/0_abstract.tex
\begin{abstract}
We present \projectname, a compact active hyperspectral video system designed for real-time capture of dynamic scenes.
Our approach combines a narrowband LED array with a coded-exposure-pixel (CEP) camera capable of high-speed, per-pixel exposure control, enabling joint encoding of scene information across space, time, and wavelength within each video frame. 
Unlike passive snapshot systems that divide light across multiple spectral channels simultaneously and assume no motion during a frame's exposure, \projectname actively synchronizes illumination and pixel-wise exposure, improving photon utilization and preserving spectral fidelity under motion. 
A learning-based reconstruction pipeline then recovers 31-channel hyperspectral (400–700\,nm) video at 30\,fps and VGA resolution, producing temporally coherent and spectrally accurate reconstructions.
Experiments on synthetic and real data demonstrate that \projectname significantly improves reconstruction fidelity and temporal stability over existing snapshot hyperspectral imaging systems, enabling robust hyperspectral video across diverse materials and motion conditions.
\end{abstract}

%% file: sections/1_introduction.tex
\vspace{-1em}
\section{Introduction}
\label{sec:intro}

Hyperspectral imaging (HSI) captures scene reflectance across many contiguous wavelength bands, revealing spectral features invisible to conventional RGB cameras. 
This rich spectral information underpins a wide range of applications such as material classification~\cite{zhi2019multispectral}, physiological monitoring~\cite{lu2014medical}, and spectral relighting~\cite{legendre2016practical}.
Despite decades of progress, delivering HSI at sufficiently high speeds to enable \textit{hyperspectral video} remains a fundamental challenge due to trade-offs between spectral resolution, light efficiency, and temporal sampling.

Traditional HSI systems rely on spatial or spectral scanning, in which different bands are sequentially captured using tunable filters, gratings, or moving optics. 
These approaches provide high spectral fidelity but require long acquisition times, making them impractical for dynamic scenes. 
Snapshot HSI methods, such as those using coded apertures (e.g., CASSI)~\cite{arce2013compressive}, diffractive optical elements (DOEs)~\cite{baek2021single}, or multispectral filter arrays (MSFAs)~\cite{geelen2014compact}, address this limitation by compressing spectral information into a single exposure through static optical encodings. 
Still, these designs suffer from severe light loss and ill-posed inversions that amplify noise and motion artifacts. 
Such issues are particularly salient when capturing hyperspectral video.
Short exposure times result in fewer photons per band, and scene motion causes smearing or ghosting due to temporally misaligned spectral samples.

One way to improve optical throughput is to shift spectral modulation from passive optics that attenuate or disperse light to the illumination itself. 
In this paradigm, the emitted spectrum is actively varied in time and/or space to encode wavelength information directly in the light rather than through optical filtering. 
Active hyperspectral systems have explored this principle using temporally multiplexed narrowband light sources~\cite{park2007multispectral, verma2024chromaflash} or spectrally structured projection~\cite{shin2024dense, li2019pro}, offering programmable control over when and where different spectral bands illuminate the scene. 
However, most of these systems achieve fine control along only a single dimension (e.g., high-speed spectral modulation with LEDs or spatial selectivity with projectors).

In a system called \textit{\projectname} (see~\autoref{fig:teaser}), we demonstrate that motion-robust hyperspectral video can be achieved by jointly encoding space, time, and spectra within each video frame. 
\projectname combines a narrowband LED array and a cutting-edge \textit{coded-exposure-pixel} (CEP) image sensor~\cite{wei2018coded, luo2017exposure, zhang2016compact, gulve202339}. 
As the system rapidly cycles through multiple LED activations, pixels are exposed in a time-varying mosaic pattern to create a dense \textit{spatio-spectro-temporal encoding} of the scene within a single video frame.
\projectname's hardware acts as an optical encoder, while a software pipeline decodes the signal through spectral demosaicing, motion alignment via optical flow, and a learning-based reconstruction stage that recovers 31-channel hyperspectral video (400-700\,nm) from coded video frames captured at 30\,fps.
Unlike prior systems that relied on bulky, alignment-sensitive, and aberration-prone optics~\cite{liu2013efficient, vargas2021time, feng2016per}, our signal acquisition operates entirely in silicon, enabling compact and calibration-free deployment.
Rather than filtering broadband lighting, \projectname's active approach allows each LED's narrowband output to contribute fully to the captured signal, thereby improving photon efficiency under dynamic or low-light conditions.

\vspace{0.5em}
\noindent
\textbf{Our main technical contributions include:}
\begin{s_itemize}
    \item \projectname, a novel hyperspectral video system that leverages time-varying illumination with pixel-wise coded-exposure to densely encode information across space, time, and wavelength.
    
    \item A compact hardware prototype that integrates a CEP sensor with a narrowband LED array capable of modulating light at microsecond scales, enabling real-time capture of dynamic spectral phenomena at 30~fps. 
    
    \item A jointly designed illumination-exposure coding scheme and reconstruction pipeline that estimates accurate and temporally coherent hyperspectral video with 31 channels spanning 400-700\,nm at VGA resolution ($640 \times 480$).

\end{s_itemize}

\noindent
Through extensive experiments on both synthetic and real data, we demonstrate that \projectname achieves significant improvements in reconstruction accuracy over state-of-the-art snapshot HSI approaches. We showcase its performance across diverse scenes with varying spatial, spectral, and motion characteristics, marking a step toward general-purpose, motion-robust hyperspectral videography.

%% file: sections/2_related_work.tex
\vspace{-1em}
\section{Related Work}
\label{sec:related}

\noindent
\textbf{Snapshot Hyperspectral Imaging.}
Efforts to enable snapshot HSI have primarily focused on encoding spectral information optically in hardware, yielding compressed measurements that can be expanded via computational reconstruction. 
Early approaches such as CASSI~\cite{arce2013compressive} employed dispersive optics, coded apertures, and relay lenses to achieve single-shot acquisition but at the expense of bulky equipment, extensive calibration, and complex inversion.
Subsequent designs improved compactness yet introduced new limitations: prism-based systems~\cite{baek2017compact} suffer from chromatic distortion and spatial blur due to dispersion; printed color-dot masks~\cite{zhao2019hyperspectral} have low optical throughput and ill-conditioned spectral responses; and MSFAs~\cite{geelen2014compact} capture very little light due to absorptive filtering, attenuating effective signal contribution under time-constrained scenarios such as motion and video. 
DOEs~\cite{baek2021single, shi2024learned, li2022qdo} and metasurface array approaches~\cite{makarenko2022real, faraji2019hyperspectral} offer high spectral diversity in miniature form factors but remain sensitive to fabrication imperfections, wavelength-dependent efficiency, and static spatial encoding that breaks under high-speed motion or noise.
Without explicit temporal coding, any scene movement during exposure introduces spatial–spectral mixing that is difficult to disentangle during reconstruction.
Recent work from Shi et al.~\cite{shi2024learned} reports using a 400\,ms exposure time under indoor illumination conditions. 
At such timescales, these systems are suitable for static imaging, yet fundamentally ill-suited for dynamic or motion-robust hyperspectral video.

\vspace{0.5em}
\noindent
\textbf{Active Hyperspectral Imaging.}
In contrast, active HSI approaches leverage programmable light sources to modulate spectral content temporally or spatially. 
This often improves the effective contribution of incoming light to useful signal---an advantage particularly beneficial under time-constrained conditions such as hyperspectral video.
A common strategy involves sequentially cycling narrowband LEDs or laser diodes~\cite{park2007multispectral, goel2015hypercam, verma2024chromaflash}, synchronizing them with the sensor to interleave spectral measurements across time.
While conceptually simple, this strategy struggles in dynamic scenes where motion between frames causes spectral misalignment and inconsistent reflectance recovery.
Verma et al.~\cite{verma2024chromaflash} mitigated this limitation by exploiting a rolling-shutter camera to multiplex spectral information within a single exposure, but as with all row-wise schemes, fast motion still produces characteristic rolling-shutter distortions.

Structured-light methods extend this idea by projecting spatially-coded spectral patterns onto the scene~\cite{shin2024dense, shin2024dispersed, li2022deep, li2019pro}, enabling sparse sampling of spatio-spectral content across multiple illumination patterns.
However, the need for multiple projection patterns and exposures to span the full spectral range restricts capture speed.
For instance, Shin et al.~\cite{shin2024dense} presented a system that achieved video capture at only 6.6~fps, which is inadequate for fast motion.
More recently, Yu et al.~\cite{yu2025active} combined a high-speed event camera with a synchronized “sweeping-rainbow” illumination pattern, encoding spectral changes as asynchronous temporal events to achieve high temporal resolution.
However, their reliance on mechanically rotating optics limits compactness and robustness, and the moving spectral sweep introduces aliasing and misalignment for motion orthogonal to the scan.

\vspace{0.5em}
\noindent
\textbf{Coded-Exposure-Pixel Imaging.}
A new class of programmable sensors known as multi-tap, multi-bucket, or coded-exposure-pixel sensors~\cite{wei2018coded, luo2017exposure, kang2022indirect, bub2010temporal, gulve202339} pushes the boundaries of conventional imaging by enabling per-pixel exposure modulation directly in hardware, allowing dense spatio-temporal coding without external optical modulation.
Modern implementations of this sensor architecture achieve VGA spatial resolution with pixel-wise modulation rates exceeding 39 kHz~\cite{gulve202339}, unlocking unprecedented flexibility for encoding scene dynamics.
While coded-exposure imaging in general has been explored for motion deblurring~\cite{raskar2006coded}, HDR imaging~\cite{martel2020neural}, and transient analysis~\cite{liu2013efficient}, its potential for hyperspectral video remains largely untapped.
The ability to embed high-speed temporal coding directly at the pixel level presents a unique opportunity to pair with programmable illumination for joint spectral–temporal multiplexing.
Leveraging this capability, \projectname introduces a tightly integrated coding scheme where time-varying spectral illumination and pixel-wise exposures are jointly tiled across the sensor in a dense mosaic.
Combined with a learning-based reconstruction pipeline, \projectname achieves real-time capture at 30\,fps and motion-robust recovery of hyperspectral video, preserving both spatial detail and spectral fidelity.

%% file: sections/3_image_formation.tex
\section{Imaging Forward Model}
\label{sec:image-formation}
\projectname combines time-varying illumination with spatio-temporal exposure modulation to achieve dense spatio-spectro-temporal coding within each video frame. 
This section describes the imaging principles, coding scheme, and measurement model underlying the system.

\subsection{Coded-Exposure-Pixel Camera}

\begin{figure}[t]
    \centering
    \includegraphics[width=\linewidth]{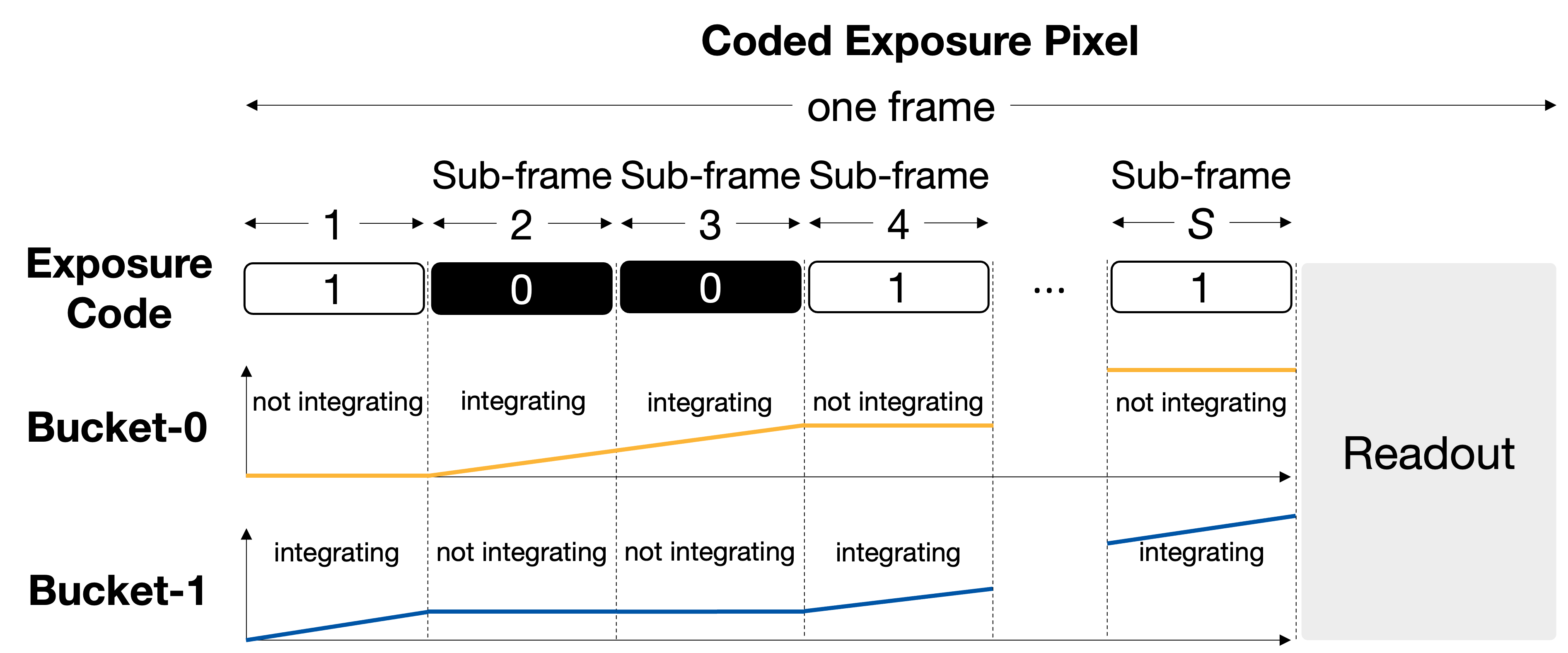}
    \caption{
        The operation of a single pixel in a coded-exposure-pixel (CEP) camera. Each pixel alternates between two charge storage sites (Bucket~0 and Bucket~1) according to a binary exposure code. During each sub-frame, only one bucket integrates incident light, enabling temporal multiplexing across the exposure period. 
    }
    \label{fig:coded_exposure_pixel}
    \vspace{-1em}
\end{figure}

A CEP camera differs from a conventional CMOS image sensor in two key features. 
First, each pixel of the CEP camera incorporates two charge-collection sites, or \textit{buckets}, enabling the segregation of photo-induced charge accumulation. 
Second, the exposure duration of each pixel within a video frame is divided into smaller intervals called \textit{sub-frames}. 
Each pixel includes a one-bit writable memory that facilitates programmable control over which bucket is active during each sub-frame.

\autoref{fig:coded_exposure_pixel} illustrates the operation of a CEP camera pixel and the corresponding integration process over sub-frames.
Charge accumulation occurs at two different timescales: 
\begin{s_enumerate}
\item \textbf{Sub-frame level:} The charge collected at each sub-frame is accumulated and transferred to the active bucket based on each pixel's control signal, and 
\item \textbf{Frame level:} The total charge accumulated across all sub-frames is integrated and read out for each bucket. 
\end{s_enumerate}

\noindent
Programming a CEP camera involves specifying the number of sub-frames per video frame, the duration of each sub-frame, and the state of each pixel's exposure at each sub-frame. 
The exposure schedule is represented by a binary matrix \( \mathbf{C} \in \{0,1\}^{P \times S} \), where \( P \) is the number of pixels and \( S \) is the number of sub-frames per frame. Each entry \( \mathbf{C}_{p,s} \) specifies which bucket is active for a given pixel $p$ and sub-frame $s$. 

\subsection{Time-Varying Spectral Illumination}
We deliver spectral modulation by using an array of \( L \) narrowband LEDs, each with a distinct spectral power distribution $E_l(\lambda)$, where $\lambda$ denotes wavelength.
The illumination is varied in synchrony with the sub-frame exposure modulation of the CEP camera.

We define the illumination schedule using a binary matrix \( \mathbf{I} \in \{0,1\}^{S \times L} \), where \( {I}_{s,l} = 1 \) indicates that LED \( l \) is active during sub-frame \( s \). 
The $s$-th row \( \mathbf{i}_{s,:} \in \{0,1\}^L \) thus specifies which LEDs are activated during the given sub-frame.
The resulting spectral irradiance incident on the scene for a given sub-frame is
\begin{equation}
    \mathcal{I}_{s}(\lambda) = \sum_{l=1}^L {I}_{s,l} \cdot E_l(\lambda).
    \label{eqn:time-var-illum}
\end{equation}
This temporally varying illumination encodes spectral content directly into the exposure sequence.
Together, the illumination matrix \( \mathbf{I} \) and the exposure code matrix \( \mathbf{C} \) form the basis for our measurement model.

\subsection{Joint Illumination and Exposure Coding}

\begin{figure*}[t]
    \centering
    \includegraphics[width=\textwidth]{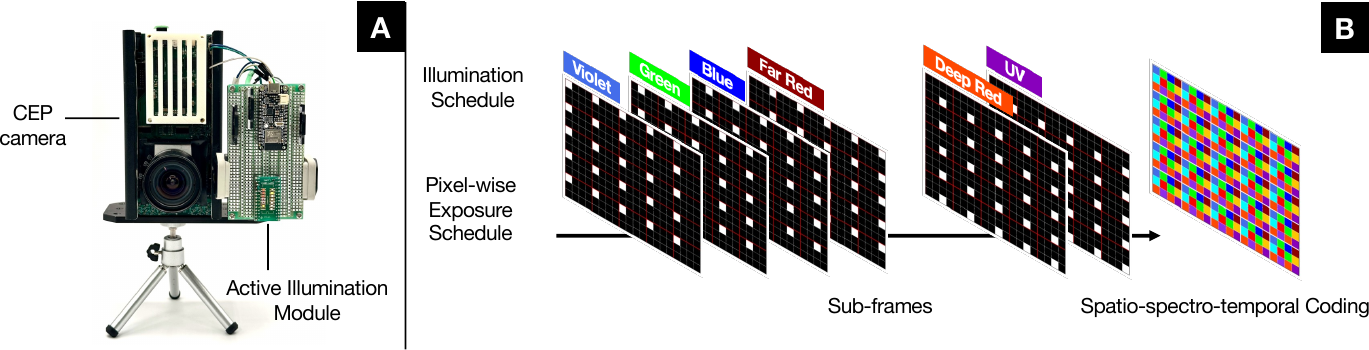}
    \caption{
        \textbf{(A)} \projectname features a CEP camera synchronized with a programmable array of narrowband LEDs. 
        \textbf{(B)} Each sub-frame is assigned a unique LED and spatial exposure mask, producing a dense spatio-spectro-temporal scene encoding within a single video frame.}
    \label{fig:illumination_exposure_coding}
    \vspace{-0.2in}
\end{figure*}

To enable dense and structured multiplexing, we design a coding scheme that synchronizes LED activations with pixel exposures across tightly packed neighborhoods on the sensor.
As shown in \autoref{fig:illumination_exposure_coding}\,(B), we cycle through \( L \) narrowband LEDs across \( S \) sub-frames while spatially tiling both illumination and exposure schedules across the sensor in a mosaic-like fashion. 
Pixels are grouped into $T$ repeating tiles (e.g., a $4 \times 4$ pixel mosaic yields \( T = 16 \) tiles) with unique exposure schedules.

We define tile exposure and illumination codes as
\begin{equation}
\mathbf{C}_{\text{tile}} \in \{0,1\}^{T \times S}, \quad 
\mathbf{I}_{\text{tile}} \in \{0,1\}^{T \times S \times L}.
\end{equation}
Here, $\mathbf{C}_{\text{tile}}[t,s] = 1$ indicates that pixels in tile $t$ are actively integrating light during sub-frame $s$, and $\mathbf{I}_{\text{tile}}[t,s,l] = 1$ specifies that LED $l$ is turned on during sub-frame $s$ for tile $t$.
We define a mapping function 
\begin{equation}
\pi: \{1, \dots, P\} \rightarrow \{1, \dots, T\}
\end{equation}
that assigns each pixel index $p$ to its corresponding tile index $\pi(p)$. 
Using this mapping, each pixel inherits its exposure and illumination codes as
\begin{equation}
{C}_{p,s} = \mathbf{C}_\text{tile}[\pi(p), s], \quad {I}_{p,s,l} = \mathbf{I}_\text{tile}[\pi(p), s, l].
\end{equation}
Thus, all pixels with the same tile index share the same illumination pattern, while adjacent pixels observe different wavelength bands at different times. 
Over $S$ sub-frames, this coordinated scheme produces a temporally staggered \textit{spectral mosaic}—a single video frame in which different spatial locations encode distinct spectral and temporal samples of the scene.

\subsection{Measurement Model}
The goal of our measurement model is to represent each hyperspectral video frame as a matrix \( \mathbf{R} \in \mathbb{R}^{P \times \Lambda} \), where each row \( \mathbf{r}_p \in \mathbb{R}^\Lambda \) describes the spectral reflectance at a pixel \( p \) according to \( \Lambda \) discrete wavelength bands.

From \autoref{eqn:time-var-illum}, the time-varying illumination at sub-frame $s$ and pixel $p$ can be expressed in discrete form as
\begin{equation}
\boldsymbol{\mathcal{I}}_{p,s} = \sum_{l=1}^{L} I_{p,s,l}\,\mathbf{E}_l,
\label{eqn:illum_discrete}
\end{equation}
where $\mathbf{E}_l \in \mathbb{R}^{\Lambda}$ denotes the spectral power distribution of LED $l$, and $I_{p,s,l} \in \{0,1\}$ specifies whether that LED is active for pixel $p$ during sub-frame $s$ according to its tile schedule.
Given the camera's spectral sensitivity $\mathcal{S} \in \mathbb{R}^{\Lambda}$, the effective spectral sensing vector is
\begin{equation}
\mathbf{a}_{p,s} = \mathcal{S} \odot \boldsymbol{\mathcal{I}}_{p,s},
\label{eqn:effective_vector}
\end{equation}
where $\odot$ denotes element-wise multiplication across the $\Lambda$ wavelength bands.  
The photo-response for pixel $p$ in sub-frame $s$ is then
\begin{equation}
y_{p,s} = \mathbf{a}_{p,s}^{\top}\mathbf{r}_p = 
\sum_{k=1}^{\Lambda} \mathcal{S}_k\,\boldsymbol{\mathcal{I}}_{p,s,k}\,\mathbf{r}_{p,k}.
\label{eqn:subframe_response}
\end{equation}
Although the CEP camera supports dual-bucket readout, we discard measurements from Bucket~0 and retain only the integration from Bucket~1 in our implementation.
The total measured intensity at pixel $p$ over a frame is obtained by summing all sub-frames in which it is active:
\begin{equation}
Y_p = \sum_{s=1}^{S} C_{p,s}\,y_{p,s} + \eta_p,
\label{eqn:pixel_measurement}
\end{equation}
where $C_{p,s} \in \{0,1\}$ indicates whether pixel $p$ integrates light during sub-frame $s$, and $\eta_p$ accounts for sensor noise and other residual errors.

Stacking all pixel measurements into a vector $\mathbf{Y} \in \mathbb{R}^{P}$ and vectorizing the hyperspectral cube as $\mathbf{x} = \mathrm{vec}(\mathbf{R}) \in \mathbb{R}^{P\Lambda}$ gives the global forward model:
\begin{equation}
\mathbf{Y} = \mathbf{A}\,\mathbf{x} + \boldsymbol{\eta},
\label{eqn:global_model}
\end{equation}
where $\mathbf{A} \in \mathbb{R}^{P \times P\Lambda}$ encodes the combined effects of illumination modulation, exposure coding, and sensor spectral sensitivity.  
Each row of $\mathbf{A}$ represents the effective spectral integration weights for one pixel aggregated over all active sub-frames.  
This linear model defines how spatial, spectral, and temporal information are jointly encoded in a single captured frame, forming the basis for our hyperspectral reconstruction pipeline.

\subsection{Hardware Prototype}
\projectname consists of a custom active illumination module and a prototype CEP camera (see~\autoref{fig:illumination_exposure_coding}\,(A)).

\vspace{0.5em}
\noindent
\textbf{Active Illumination Module.}
The illumination module comprises 12 high-power narrowband LEDs (Lumileds Luxeon C), each covering a distinct portion of the visible spectrum with a full-width-at-half-maximum (FWHM) of approximately 20–30\,nm.
All LEDs are driven by a custom current driver capable of switching at frequencies exceeding 100\,kHz.
A microcontroller (Adafruit ESP32 Feather v2) generates digital control signals that synchronize LED activation with the CEP camera's sub-frame clock, achieving microsecond-level timing precision.

\vspace{0.5em}
\noindent
\textbf{CEP Camera.}
The prototype uses a VGA-resolution ($640 \times 480$) CEP image sensor that supports per-pixel binary exposure modulation.
The sensor operates at up to 12,500 sub-frames per second, allowing rapid alternation of illumination states within a single video frame.
In our implementation, each video frame comprises $S=158$ sub-frames of 170\,µs each, yielding a total integration period of $\sim$27\,ms that is suitable for 30-fps video.
An additional $\sim$6\,ms of readout and synchronization overhead introduces dead time between consecutive frames, making this the primary factor limiting the achievable frame rate to 30\,fps rather than the exposure duty cycle itself.

\vspace{0.5em}
\noindent
\textbf{Synchronization and Control.}
LED activation and sensor exposure schedules are jointly programmed and triggered via hardware-level synchronization lines.
Each LED's activation window is mapped to a contiguous sequence of sub-frames that share a common exposure pattern.
To compensate for variations in LED output power and the camera's spectral sensitivity, LEDs with lower radiance are allocated proportionally longer activation durations, ensuring balanced spectral energy delivery across the frame.
No two LEDs are active simultaneously to minimize crosstalk and improve spectral separability.

\vspace{0.5em}
\noindent
\textbf{Calibration.}
We perform a one-time calibration procedure to characterize (1) the spectral power distribution of the illumination module, (2) the spectral sensitivity of the CEP camera, and (3) the system's overall radiometric response. 
These measurements define the parameters of the imaging forward model while ensuring that reconstructed hyperspectral data is both radiometrically accurate and physically consistent with real-world measurements.

Each of the 12 LEDs is sequentially activated so that its emission spectrum $\mathbf{E}_l$ can be measured with a calibrated spectroradiometer (Konica Minolta CS-2000), providing an accurate spectral power distribution from 380–780\,nm. 
The camera's spectral sensitivity $\mathcal{S}$ is measured using a monochromator (Image Engineering camSPECS XL) with 5-nm interference filters; the outputs are aggregated into 10-nm bins for alignment with the reconstruction model.
To ensure accurate radiometric scaling, we capture a Macbeth ColorChecker and jointly optimize LED-specific gains to equalize integrated intensity across channels, compensating for residual non-uniformities in LED brightness, optical coupling, and sensor response.
We use the resulting calibration parameters in both simulation and real-world reconstruction. 
Detailed procedures and plots are provided in Supplementary~\autoref{supp:system}.

%% file: sections/4_reconstruction.tex
\vspace{-1em}
\section{Hyperspectral Video Reconstruction}
\label{sec:reconstruction}
Given a coded measurement $\mathbf{Y}_{i}$ corresponding to frame $i$, our goal is to reconstruct the underlying hyperspectral scene $\mathbf{R}_{i} \in \mathbb{R}^{P \times \Lambda}$, where each pixel's spectral reflectance $\mathbf{r}_{p}$ spans $\Lambda=31$ wavelength bands. 
We first demosaic $\mathbf{Y}_{i}$ into a set of LED-specific sub-images $\{\mathbf{Y}^{(1)}_{i}, \mathbf{Y}^{(2)}_{i}, \dots, \mathbf{Y}^{(L)}_{i}\}$, 
where $L=12$ denotes the number of narrowband LEDs. 
Each $\mathbf{Y}^{(l)}_{i}$ aggregates pixels assigned to LED $l$ according to the known illumination–exposure mosaic described in \autoref{sec:image-formation}. 
We then perform bilinear interpolation to upsample each sub-image to the full spatial resolution of the sensor.

Unlike with passive filter array systems, our sub-images correspond to different time intervals within the exposure period, meaning they may be spatially misaligned in scenes with motion.
To avoid artifacts during reconstruction, we incorporate a temporal alignment step beforehand.

\subsection{Temporal Alignment of Spectral Channels}
Each LED-specific sub-image $\mathbf{Y}_{i}^{(l)}$ represents scene content captured under a distinct narrowband illumination. 
This fact violates the photometric consistency assumption underlying conventional optical flow methods, making direct alignment between sub-images unreliable.
To address this, we perform temporal alignment by estimating motion across sub-images corresponding to the same LED in adjacent frames.

We designate the sub-image illuminated by the lime-coloured LED, $\mathbf{Y}_{i}^{(\text{lime})}$, as the temporal reference for frame~$i$ due to its central wavelength and mid-exposure timing. 
For each other LED $l$, we determine its relative position in the illumination schedule and pair its sub-image $\mathbf{Y}_{i}^{(l)}$ with a temporally adjacent frame. 
If $l$ occurs before the lime LED in the illumination cycle, we pair it with $\mathbf{Y}_{i+1}^{(l)}$; if it occurs after, we pair it with $\mathbf{Y}_{i-1}^{(l)}$. 
Using these pairs, we estimate motion fields with the Real-Time Intermediate Flow Estimation (RIFE) network~\cite{huang2022rife}, which predicts an interpolated frame at a specified timestamp. 
We warp each sub-image $\mathbf{Y}_{i}^{(l)}$ to the reference time of $\mathbf{Y}_{i}^{(\text{lime})}$, producing the temporally aligned result $\hat{\mathbf{Y}}_{i}^{(l)}$.

\subsection{Learning-based Reconstruction}
\label{sec:learning-reconstruction}
For each video frame $i$, the temporally aligned sub-images 
$\{\hat{\mathbf{Y}}^{(1)}_{i}, \hat{\mathbf{Y}}^{(2)}_{i}, \dots, \hat{\mathbf{Y}}^{(L)}_{i}\}$ 
serve as input to a deep neural network that reconstructs the corresponding hyperspectral image 
$\mathbf{R}_{i}$ with $\Lambda = 31$ spectral channels.

\vspace{0.5em}
\noindent
\textbf{Model Architecture.}
We adopt the Holistic Attention Network (HAN)~\cite{niu2020han} as our reconstruction backbone because of its demonstrated effectiveness in image restoration tasks. 
The network consists of 18 residual blocks organized into 10 residual groups, each with 128 feature channels. 
Channel attention with a reduction ratio of 16 is applied to enhance feature discrimination across spectral bands.

The model takes as input a $66 \times 64 \times 12$ tensor, corresponding to a spatial crop of the 12 demosaiced and temporally aligned LED sub-images.
The model outputs a $66 \times 64 \times 33$ hyperspectral cube covering 33 spectral bands from 380–780\,nm as follows:
\begin{s_itemize}
    \item \textbf{Channel 1:} Aggregated ultraviolet bands (380–390\,nm),
    \item \textbf{Channels 2–32:} Consecutive 10\,nm bins spanning the visible range (400–700\,nm),
    \item \textbf{Channel 33:} Aggregated near-infrared bands (710–780\,nm).
\end{s_itemize}

\noindent
Extrapolated edge channels (1 and 33) are included during training to improve reconstruction near the spectral boundaries but are discarded at inference, leaving Channels 2–32 as the final output. 
We reconstruct hyperspectral images corresponding to full frames in a patch-wise manner and merge them using a weighted aggregation strategy to ensure seamless spatial continuity.  
Additional details on model architecture, patch-wise reconstruction, and aggregation are provided in Supplementary~\autoref{supp:reconstruction}.

\vspace{0.5em}
\noindent
\textbf{Training Setup.}
To train the reconstruction model, we simulate the proposed imaging system's forward model using hyperspectral image datasets while incorporating sensor noise to closely emulate real capture conditions. 
We add zero-mean Gaussian noise with a standard deviation uniformly sampled between 0\% and 15\% of the maximum signal intensity.

For data augmentation, we extract random spatial patches from hyperspectral images and apply random horizontal and vertical flips. 
Since most public datasets provide spectral measurements only between 400–700\,nm, we extend the range to 380–780\,nm by mirroring the edge channels: the 420-nm and 410-nm bands approximate the ultraviolet (380–390\,nm) region, while channels beyond 710\,nm are mirrored from the 700-nm band. 
These extrapolated channels are used only during training to stabilize spectral boundary reconstruction and are omitted at inference.

We minimize the $\mathcal{L}_1$ loss between the predicted and ground-truth hyperspectral cubes using the Adam optimizer with a learning rate of $1\times10^{-4}$ and default $\beta$ parameters. 
We implement the model in PyTorch and train it with a batch size of 14 and gradient accumulation over two steps for memory efficiency. 
Training for 50{,}000 iterations on an NVIDIA RTX A6000 GPU takes approximately 24 hours, and inference on a single $640 \times 480$ frame requires 4.7\,s.

%% file: sections/5_evaluation.tex
\section{Experiments}
\label{sec:evaluation}
We evaluate \projectname across simulations and real-world captures, progressing from controlled static reconstructions to dynamic video demonstrations.  

\subsection{Simulations}
\label{sec:simulations}

\noindent
\textbf{Setup.} 
Since publicly available hyperspectral video datasets are rare and small, our simulations focus on static scene reconstruction.
We simulate image formation under matched conditions on a unified corpus of three datasets: CAVE~\cite{yasuma2008cave} (32 indoor scenes), KAUST~\cite{li2021kaust} (409 indoor and outdoor scenes), and ARAD~\cite{arad2022recovery} (949 indoor and outdoor scenes).
After resampling each hyperspectral cube to 31 uniformly-spaced channels between 400–700\,nm at 10-nm intervals, we divide the corpus into 80\%-10\%-10\% splits for training, validation, and testing, respectively.

We compare reconstruction performance across the five configurations: (1) \projectname's forward model with HAN~\cite{niu2020han} as its reconstruction backbone; (2) \projectname with a MCAN~\cite{feng2021mosaic} backbone; (3) \projectname with a SRNet~\cite{bian2024broadband} backbone; (4) QDO~\cite{li2022qdo}, an end-to-end optimized DOE-based snapshot HSI system; and (5) MST++~\cite{cai2022mstplusplus}, a data-driven RGB-to-hyperspectral reconstruction method.  
We train all models from scratch using identical data splits, input normalization, and optimization schedules.  
We evaluate performance on the held-out test set using standard spatial and spectral quality metrics:  
peak signal-to-noise ratio (PSNR), structural similarity index (SSIM), mean absolute error (MAE), and spectral angle mapper (SAM). 
Higher PSNR/SSIM and lower MAE/SAM indicate better reconstruction quality.

\begin{figure}[t]
    \centering
    \includegraphics[width=\linewidth]{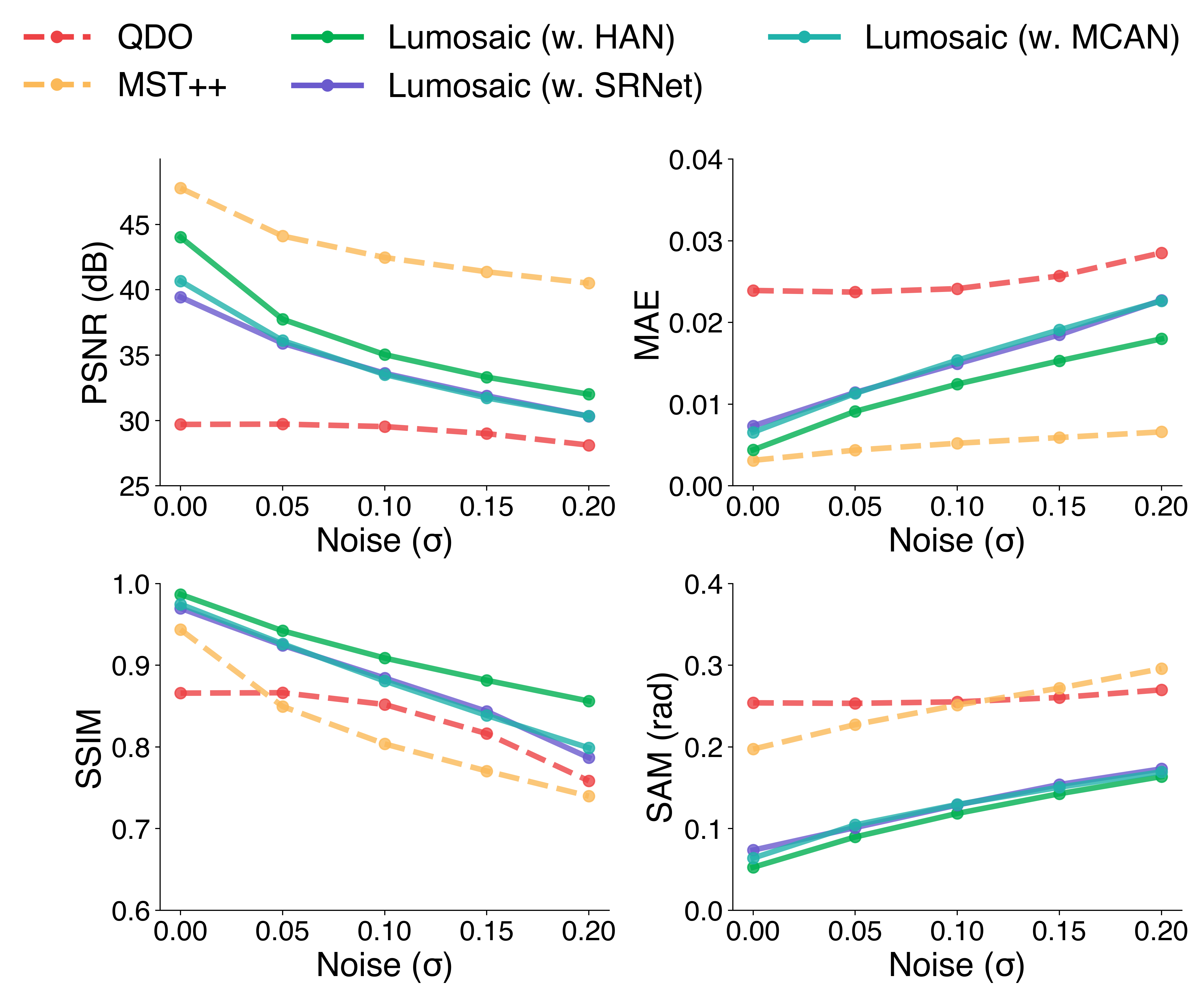}
    \caption{
    Reconstruction performance of QDO~\cite{li2022qdo}, MST++~\cite{cai2022mstplusplus}, and \projectname using HAN~\cite{niu2020han}, MCAN~\cite{feng2021mosaic}, and SRNet~\cite{bian2024broadband} backbones according to \textbf{(top-left)} PSNR, \textbf{(top-right)} MAE, \textbf{(bottom-left)} SSIM, and \textbf{(bottom-right)} SAM for the simulations under varying Gaussian noise levels. 
    }
    \label{fig:simulation-results}
\end{figure}

\vspace{0.5em}
\noindent
\textbf{Noise Robustness.}
To assess stability under photon-limited conditions typical in fast motion and video capture, we add Gaussian noise levels of $\sigma = \{0\%, 5\%, 10\%, 15\%, 20\%\}$ relative to the maximum signal intensity. 
As shown in \autoref{fig:simulation-results}, \projectname consistently achieves higher SSIM and lower SAM compared to baselines across all noise levels, while maintaining very high PSNR. 
Qualitative comparisons in Supplementary~\autoref{supp:results-simulations} show that \projectname preserves spatial details and accurate spectral reflectance, whereas QDO and MST++ exhibit noticeable blurring and spectral distortion under higher noise. 

\vspace{0.3em}
\noindent
\textbf{Reconstruction Backbone Sensitivity.}
All three \projectname-based models spectrally outperform external baselines (QDO and MST++) while maintaining high spatial fidelity, highlighting the performance gains of our sensing model rather than mere network complexity.
Among the reconstruction backbones, HAN delivers the best fidelity, reaching 44.0~dB in the noise-free case and 32.0~dB even at $\sigma = 20\%$ (see~\autoref{fig:simulation-results}).
MCAN and SRNet perform slightly worse than HAN in reconstruction accuracy but require significantly less compute and inference time (52\,ms and 27\,ms per frame, respectively), making them attractive for real-time deployment.

\vspace{0.5em}
\noindent
\textbf{Spectral Resolution.}
Supplementary~\autoref{supp:spectral-resolution} describes additional experiments on synthesized scenes to demonstrate \projectname's capability of recovering narrowband spectral features.  
Results show accurate reconstruction of sharp spectral transitions, even beyond the physical sampling limits of our 12-channel LED illumination.

\subsection{Real-World Evaluation}
\label{subsec:real-world}

\begin{figure}[t]
    \centering
    \includegraphics[width=\linewidth]{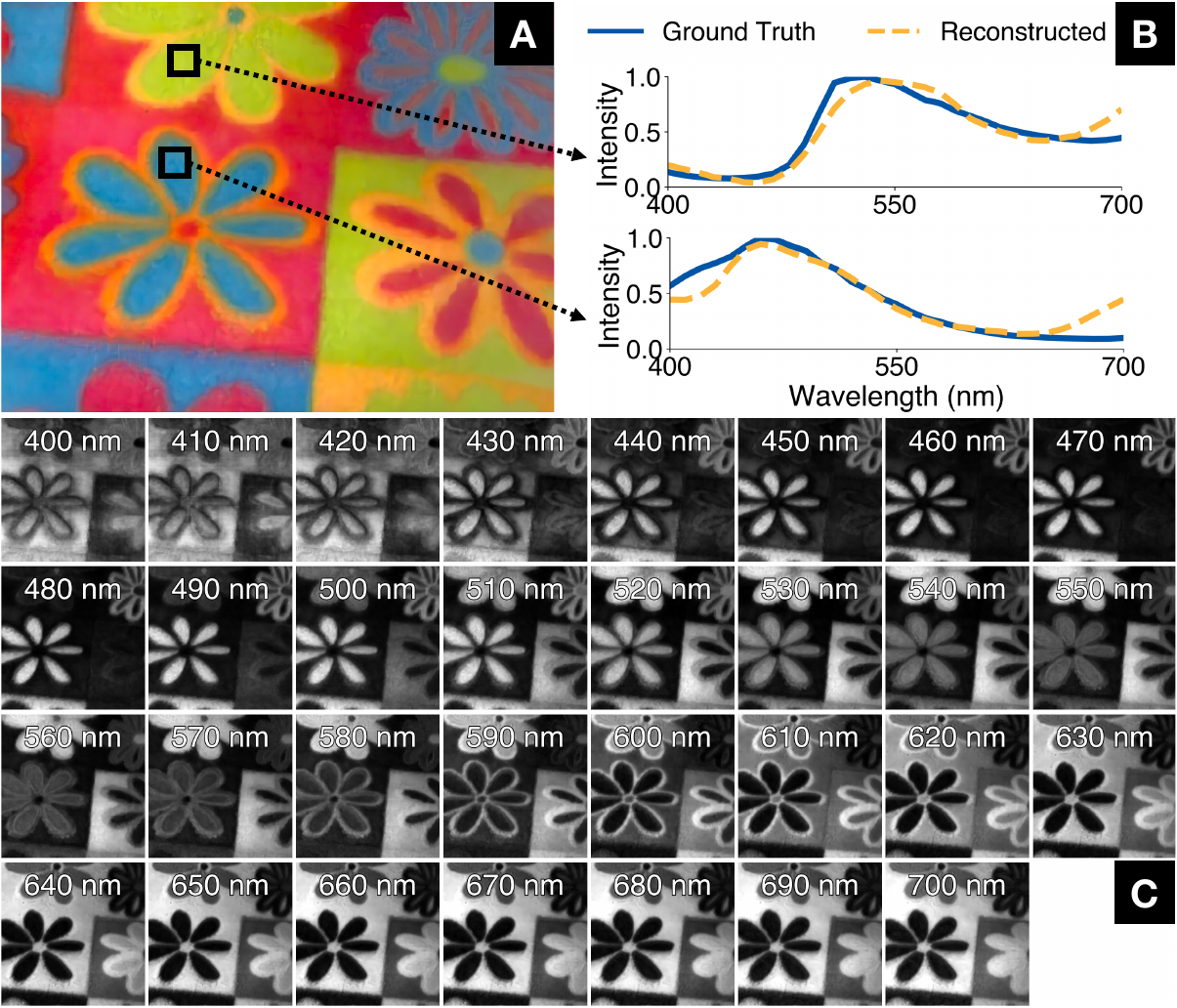}
    \caption{
        An example of real-world reconstruction results obtained using \projectname. 
        \textbf{(A)} Reconstructed hyperspectral image rendered as sRGB. 
        \textbf{(B)} Reflectance spectra extracted from marked regions. 
        \textbf{(C)} Reconstructed hyperspectral channels. 
    }
    \label{fig:static-scenes}
    \vspace{-0.2in}
\end{figure}

\vspace{0.5em}
\noindent
\textbf{Static Scenes.}
\begin{figure}[t]
    \centering
    \includegraphics[width=\linewidth]{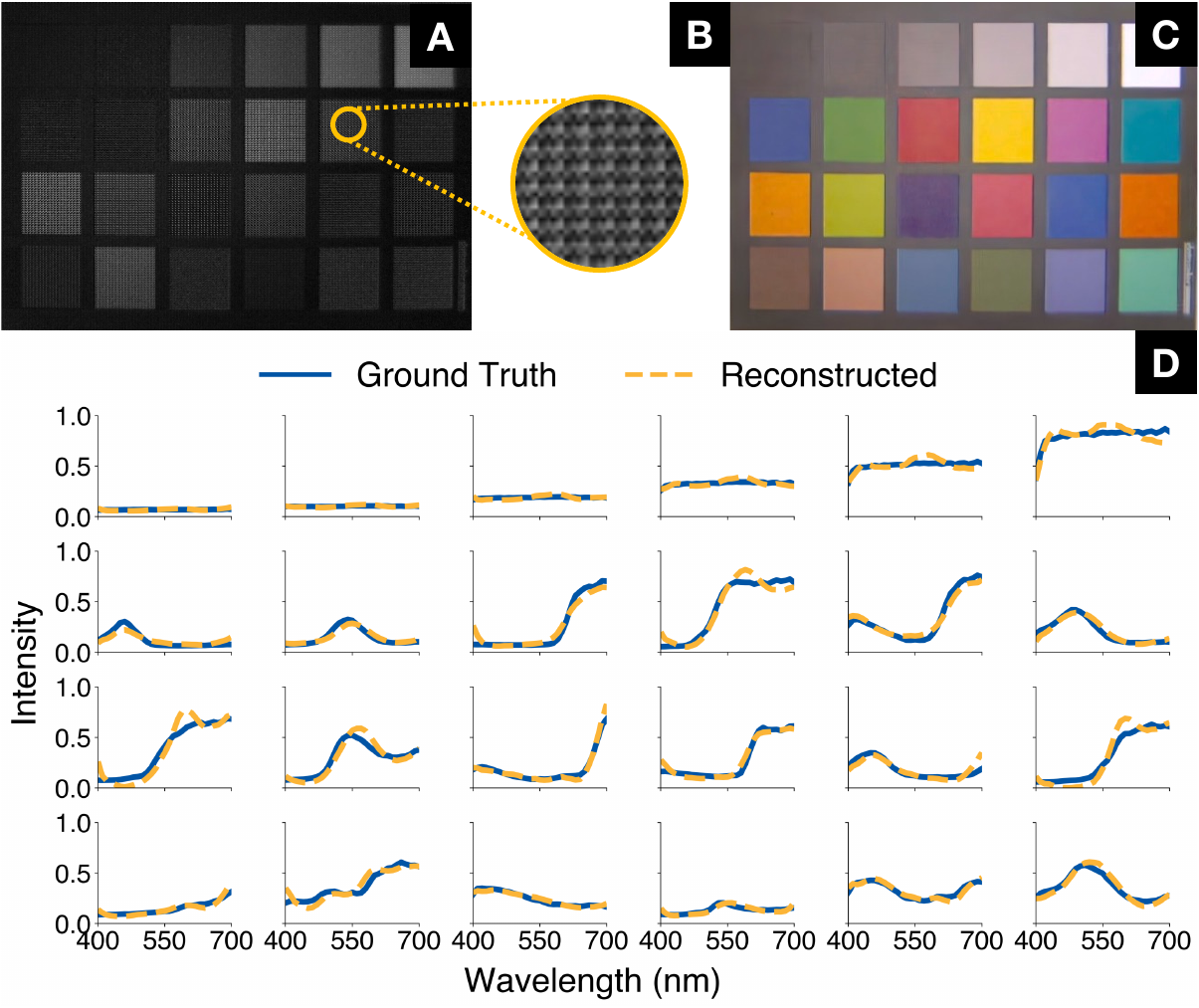}
    \caption{Quantitative and qualitative evaluation of spectral reconstruction accuracy using a ColorChecker target (real data).
    \textbf{(A)} Raw coded frame captured by \projectname. 
    \textbf{(B)} Zoomed-in region showing the mosaic coding.
    \textbf{(C)} Reconstructed hyperspectral image rendered in sRGB. 
    \textbf{(D)} Spectral reflectance curves for all 24 patches, comparing ground-truth (solid blue) and reconstructed (dashed yellow) spectra.}
    \label{fig:colorchecker}
    \vspace{-1.5em}
\end{figure}
To assess spectral and perceptual fidelity, we capture several static scenes: a ColorChecker, optical filters, and everyday objects (e.g., fabrics, printed materials, and figurines). 
As shown in \autoref{fig:colorchecker}, the reconstructed reflectance spectra from the ColorChecker closely match the ground truth from a spectroradiometer (Konica Minolta CS-2000), validating both the radiometric calibration and spectral reconstruction accuracy.  
Additional examples (\autoref{fig:static-scenes} and Supplementary~\autoref{supp:static-scenes}) show faithful color reconstruction and high perceptual consistency across diverse materials and textures.

\vspace{0.5em}
\noindent
\textbf{Metamerism Disambiguation.}
HSI can be used to discriminate between visually similar materials with different spectral properties.
We evaluate \projectname's performance at this task by imaging a genuine, pigment-based ColorChecker and its printed photocopy.
As shown in Supplementary~\autoref{fig:metamerism_results}, their reconstructed spectral reflectances differ significantly.

\begin{figure}[t]
    \centering
    \includegraphics[width=\linewidth]{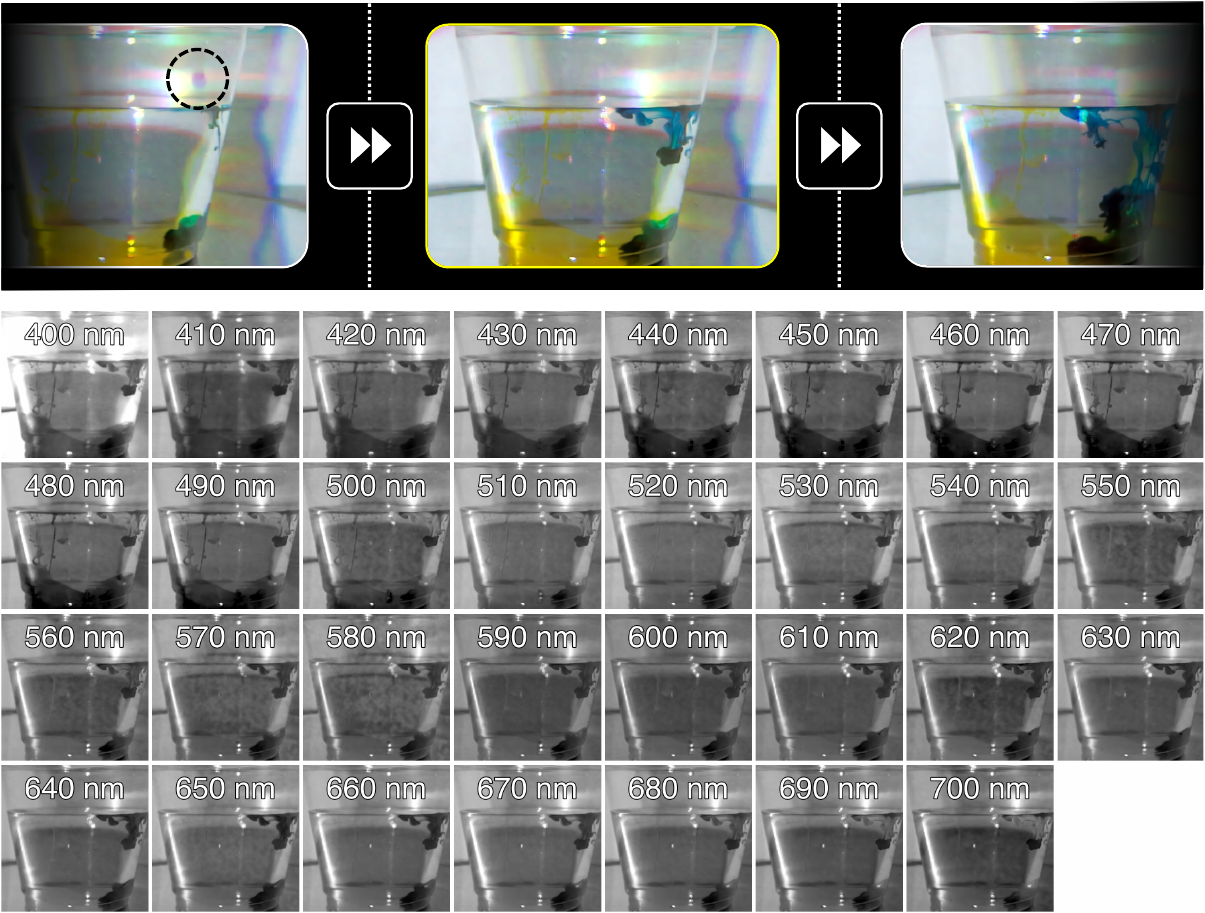}
    \caption{
        Hyperspectral video reconstruction of a dynamic scene: a colored droplet diffusing in water, captured using \projectname at 30\,fps. 
        Each column shows a representative frame over time rendered in sRGB, while the bottom row visualizes reconstructed spectral channels for the highlighted frame. 
        See the Supplementary Video for the full sequence visualization.
    }
    \label{fig:dynamic-scenes}
    \vspace{-1.5em}
\end{figure}

\vspace{0.5em}
\noindent
\textbf{Dynamic Scenes.}
Finally, we evaluate \projectname on dynamic scenes exhibiting both rigid (translation, rotation, panning) and non-rigid (hand gestures, liquid diffusion, effervescence) motions.  
Although millisecond-level offsets between LED sub-images would typically induce motion blur, \projectname reconstructs temporally coherent hyperspectral video with high spectral fidelity and stability at 30\,fps (see~\autoref{fig:teaser} and \ref{fig:dynamic-scenes}).
The Supplementary Video presents corresponding sRGB and hyperspectral renders, showing temporally stable reconstructions with minimal ghosting or flicker across all motion types.  
Additional results and ablations (Supplementary~\autoref{supp:dynamic-scenes} and \ref{supp:ablation-temporal}) highlight the contribution of our flow-based temporal alignment step in reducing motion-induced artifacts.

%% file: sections/6_conclusion.tex
\section{Future Work \& Concluding Remarks}
\label{sec:conclusion}
We have demonstrated that \projectname can reconstruct 30-fps hyperspectral video at VGA resolution with 31 spectral channels (400-700\,nm).
This was made possible by coordinating time-varying illumination with a CEP camera to generate a dense spatio-spectro-temporal encoding within each video frame.
\projectname's active, co-designed sensing strategy simultaneously acquires motion and spectral information with high light efficiency---a capability not previously realized in such a compact form factor. 
Our results show temporally coherent and spectrally faithful reconstructions across diverse materials and motion patterns, bridging the long-standing gap between snapshot imaging and true hyperspectral video.

There are numerous opportunities for future investigations.
First, our reconstruction pipeline processes each frame independently, requiring us to push the limits of snapshot HSI to an acquisition speed suitable for video frame rates.
Our main limiting factor was the dearth of comprehensive hyperspectral video datasets, which prevented us from reliably training a network that could exchange information across consecutive frames. 
In future work, we will explore simulating motion in more widely available hyperspectral image datasets to overcome this limitation.
We also did not fully leverage the CEP camera's affordances.
Our implementation used only a single bucket of the underlying sensor.
Since both buckets at each pixel integrate complementary illumination states over time, jointly modeling their responses could further improve dynamic range, light efficiency, and motion robustness.
Finally, we did not fully explore the trade-offs of different coding designs, as using adaptive or randomized mosaics may yield their own advantages.

In summary, \projectname\ establishes a new design space for computational hyperspectral video by coupling active illumination with coded-exposure imaging.
We envision this framework to enable new opportunities for real-time spectral sensing in robotics, microscopy, and computational photography.

%% file: suppl.tex
\clearpage
\setcounter{page}{1}
\setcounter{section}{0}
\setcounter{subsection}{0}
\setcounter{figure}{1}
\setcounter{table}{1}
\setcounter{equation}{0}

\renewcommand{\theHsection}{supp.\arabic{section}}
\renewcommand{\theHsubsection}{supp.\arabic{section}.\arabic{subsection}}
\renewcommand{\theHsubsubsection}{supp.\arabic{section}.\arabic{subsection}.\arabic{subsubsection}}
\renewcommand{\theHfigure}{supp.fig.\arabic{figure}}
\renewcommand{\theHtable}{supp.tab.\arabic{table}}
\renewcommand{\theHequation}{supp.eq.\arabic{equation}}

\maketitlesupplementary

\startcontents[supp]
\printcontents[supp]{}{1}{\section*{Table of Contents}}

\input{supplementary/1_imaging_system}
\input{supplementary/2_temporal_alignment}
\input{supplementary/3_reconstruction}
\input{supplementary/4_additional_results_simulations}
\input{supplementary/5_additional_results_real_world}

\stopcontents[supp]

%% file: supplementary/1_imaging_system.tex
\section{Imaging System Prototype}
\label{supp:system}

\projectname employs a custom-built imaging platform that combines actively modulated LED illumination with a coded-exposure-pixel (CEP) camera to achieve dense spatio–spectro–temporal encoding suitable for hyperspectral video capture. 
This section details the system's hardware design, control strategy, and calibration procedure.

\subsection{Active Illumination Module}
\label{supp:illumination}
\begin{figure}[t]
    \centering
    \includegraphics[width=\linewidth]{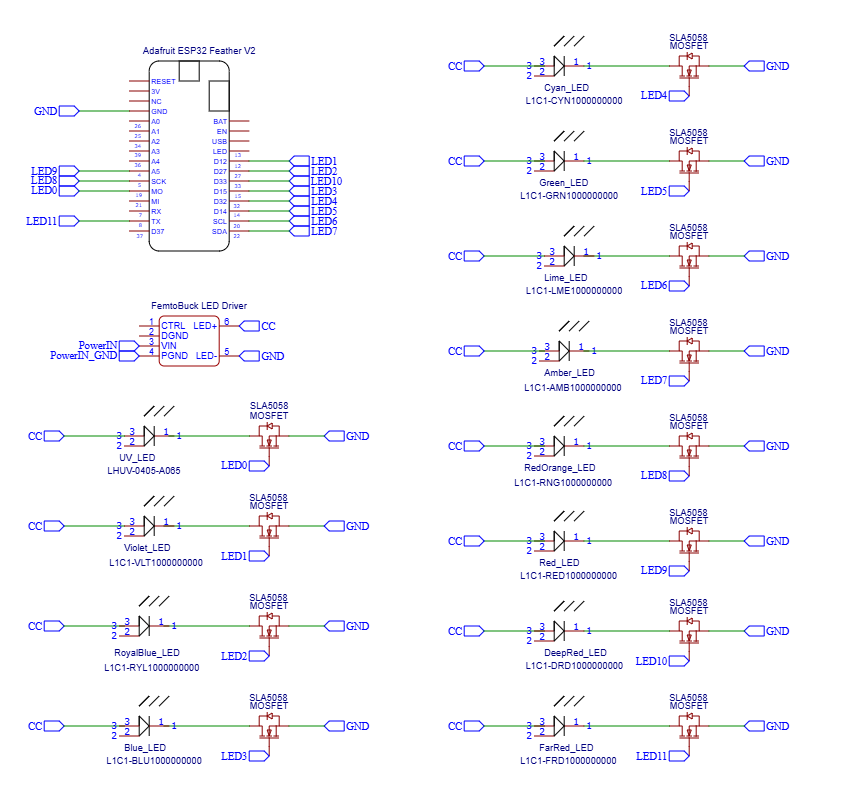}
    \caption{
        The hardware schematic of \projectname's active illumination module.
        The ESP32 microcontroller coordinates the LED driver array and issues synchronization pulses to the CEP camera.
    }
    \label{fig:hw_schem}
\end{figure}

We developed a custom high-speed illumination module to enable time-varying spectral excitation synchronized with the CEP camera (\autoref{fig:hw_schem}). 
The module integrates 12 high-power narrowband LEDs (Lumileds Luxeon C series) with full-width-at-half-maximum (FWHM) spectral bandwidths of approximately 20–30\,nm, spanning the visible range. 
The LEDs are controlled by a constant-current, high-speed switching driver capable of switching at rates exceeding 100\,kHz.

The imaging system is orchestrated by an Adafruit ESP32 Feather~v2 microcontroller, which generates digital control signals for both the illumination module and the CEP camera. 
A single microsecond-resolution clock drives all timing events, ensuring sub-frame synchronization between light modulation and sensor exposure.

In our implementation, each hyperspectral video frame consists of 158 sub-frames; each sub-frame is 150~$\mu$s in duration, corresponding to a total exposure window of 23.7~ms.
Because individual LEDs differ in radiant power and the camera exhibits wavelength-dependent sensitivity, we non-uniformly allocate sub-frame counts per LED to approximately equalize the integrated spectral energy delivered per frame. 
LEDs with lower radiance receive proportionally more sub-frames (allocated consecutively), whereas brighter ones receive fewer. 
This adaptive exposure scheduling improves both channel balance and spectral dynamic range. 
The allocation used in our experiments is summarized in \autoref{tab:led_duration}.

\begin{table}[tb]
\centering
\caption{
The LED exposure allocation per video frame compensates for relative intensity variations across LEDs and camera spectral sensitivity. Each LED's sub-frame activations occur contiguously within the frame.
}
\label{tab:led_duration}
\begin{tabular}{@{}lcc@{}}
\toprule
\textbf{LED Name} & \textbf{Relative Allocation} & \textbf{Time per Frame (\textmu s)} \\
\midrule
UV          &  5.70\% & 1,350 \\
Violet      &  3.16\% & 750  \\
Royal Blue  &  3.16\% & 750  \\
Blue        &  3.16\% & 750  \\
Cyan        &  5.70\% & 1,350  \\
Green       &  6.96\% & 1,650 \\
Lime        &  5.06\% & 1,200 \\
Amber       & 25.32\% & 6,000 \\
Red Orange  &  8.23\% & 1,950 \\
Red         &  7.59\% & 1,800 \\
Deep Red    &  6.96\% & 1,650 \\
Far Red     & 18.99\% & 4,500 \\
\midrule
\textbf{Total} & 100.00\% & 23,700 \\
\bottomrule
\end{tabular}
\end{table}


\subsection{Coding Scheme Design}
\label{supp:coding}
\begin{figure}[tb]
    \centering
    \includegraphics[width=0.5\linewidth]{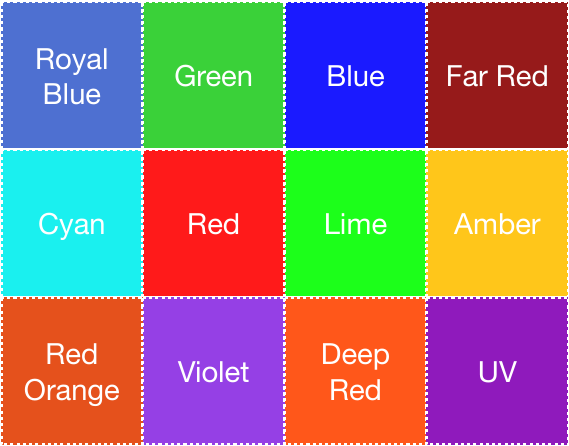}
    \caption{
    The $3\times4$ mosaic tile that forms the basis of \projectname's spatial-spectral coding.
    }
    \label{fig:mosaic}
\end{figure}

\autoref{fig:mosaic} illustrates the $3 \times 4$ spatial–spectral mosaic tile that we achieve by coordinating illumination with pixel exposures.
The pattern is executed left-to-right from the top-left corner to the bottom-right.
The mapping between LEDs and tile positions attempts to spectrally and temporally distribute LEDs with adjacent spectra in order to minimize correlations between nearby measurements.

\subsection{Calibration}
\label{supp:calibration}

To ensure accurate spectral recovery, we jointly calibrate LED intensity scaling factors and camera gain for each spectral channel. 
Calibration minimizes reconstruction error over ColorChecker measurements captured under controlled illumination.

\vspace{0.5em}
\noindent
\textbf{Measurements.}
We first measure the spectral power distribution of each LED at a fixed position using a calibrated spectroradiometer (Konica Minolta CS-2000), yielding high-resolution emission profiles $E_l(\lambda)$ from 380–780\,nm at 1-nm intervals.
We repeat this process for an incandescent bulb to generate an emission profile $B(\lambda)$.

To characterize the ColorChecker, we measure the spectral radiance of each patch $\mathcal{C}_p'(\lambda)$ under illumination from the incandescent bulb using the same spectroradiometer. 
We compute the spectral reflectance of each patch $\mathcal{C}_p(\lambda)$ by dividing the measured radiance by the incident illumination:
\begin{equation}
\mathcal{C}_p(\lambda) = \mathcal{C}'_p(\lambda) \, / \, B(\lambda).
\end{equation}

We measure the camera's spectral sensitivity $\mathcal{S}(\lambda)$ using a monochromator (Image Engineering camSPECS XL) with 39 interference filters, producing calibrated spectral response curves at a resolution of 5\,nm. 
These curves are quantized by spectral binning into $\Lambda = 41$ channels spanning 380–780\,nm in 10-nm increments, resulting in spectral vectors $\mathbf{E}_l \in \mathbb{R}^{\Lambda}$, $\mathcal{C}_p \in \mathbb{R}^{\Lambda}$, and $\mathcal{S} \in \mathbb{R}^{\Lambda}$.

We then capture five repeated measurements of the ColorChecker using \projectname and average them to reduce noise. 
The resulting image is then demosaiced to produce $L = 12$ sub-images, one for each LED. 
For each LED sub-image, we sample a region within each ColorChecker patch (total 24 patches) and compute the average intensity. 
This yields an empirical camera response matrix $\mathcal{M}^{\text{real}} \in \mathbb{R}^{L \times 24}$ where each entry represents the average response of the camera to a specific LED and ColorChecker patch.

\vspace{0.5em}
\noindent
\textbf{Computing Calibration Coefficients.} 
We determine a vector of LED scaling factors $\alpha \in \mathbb{R}^L$ by simulating a theoretical camera response matrix $\mathcal{M}^{\text{sim}}_{l, p}$ from a simplified version of our image formation model:
\begin{equation}
\mathcal{M}^{\text{sim}}_{l, p} = \sum_{\lambda=1}^{\Lambda} \alpha_l \cdot (\mathbf{E}_l)_\lambda \cdot (\mathcal{C}_p)_\lambda \cdot \mathcal{S}_{\lambda}.
\end{equation}

\noindent
To calculate $\alpha$, we minimize the non-negative least-squares reconstruction error between $\alpha$-scaled simulated intensities and our averaged measurements on the ColorChecker:
\begin{equation}
\alpha = \argmin_{\alpha \ge 0} \sum_{l, m} ([\sum_{\lambda=1}^{\Lambda} \alpha_l \cdot (\mathbf{E}_l)_\lambda \cdot (\mathcal{C}_p)_\lambda \cdot \mathcal{S}_{\lambda}  ] - \mathcal{M}^{\text{real}}_{l,p})^2.
\end{equation}

\noindent
The optimized $\alpha_l$ values are used to scale the LED emission curves, producing calibrated spectra $E'_l(\lambda) = \alpha_l E_l(\lambda)$ used in all forward modeling and reconstruction experiments.

\begin{figure}[tb]
    \centering
    \begin{minipage}{0.7\linewidth}
        \centering
        \includegraphics[width=\linewidth]{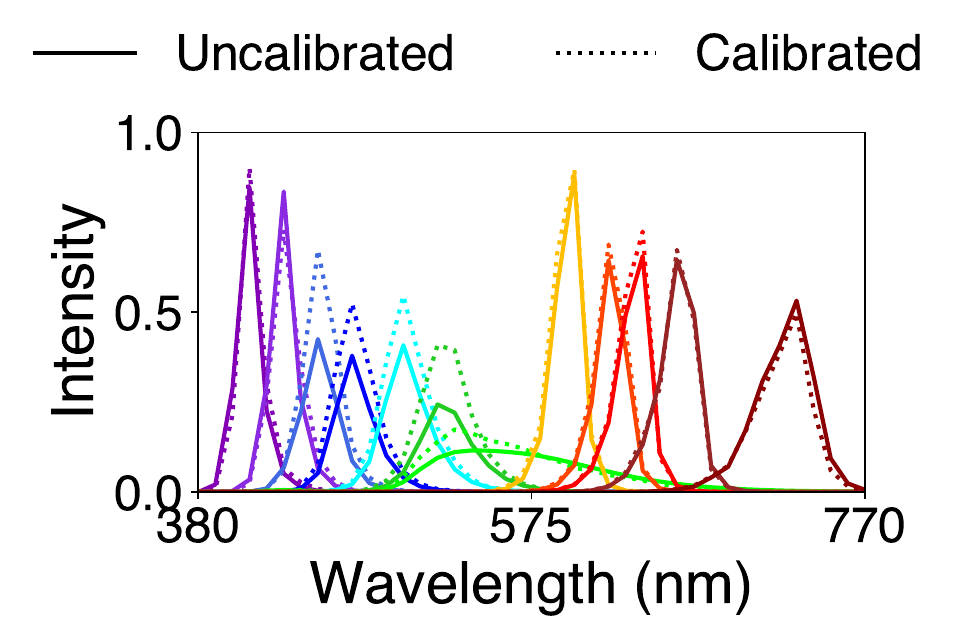}
    \end{minipage}
    \begin{minipage}{0.7\linewidth}
        \centering
        \includegraphics[width=\linewidth]{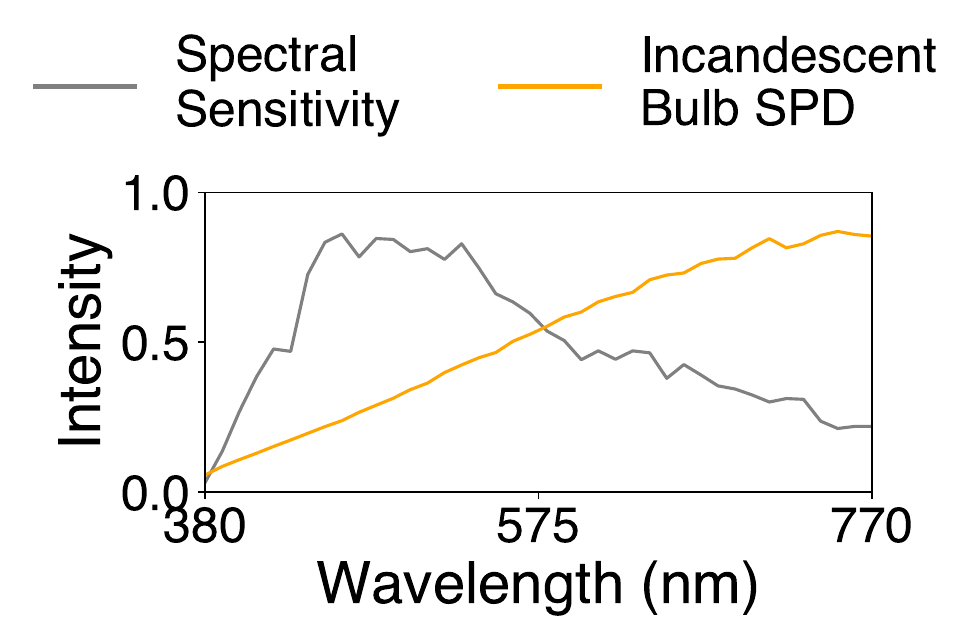}
    \end{minipage}
    \caption{\textbf{(top)} LED spectral power distributions before and after calibration via optimized scaling factors $\alpha_\lambda$. \textbf{(bottom)} Camera spectral sensitivity $\mathcal{S}(\lambda)$  and incandescent bulb spectral power distribution (SPD).}
    \label{fig:light_curves_and_sensitivity}
\end{figure}

%% file: supplementary/2_temporal_alignment.tex
\section{Temporal Alignment and Motion Compensation}
\label{supp:alignment}

Because each sub-image corresponds to a distinct time interval within the total exposure window, scenes containing motion may exhibit spatial misalignment across spectral channels. To prevent artifacts in hyperspectral reconstruction, we perform a temporal alignment step that warps all sub-images within a frame to a common temporal reference.



Each sub-image $Y^{l}_i$ in a given frame $i$ is assigned a timestamp according to its midpoint exposure timestamp.
In other words, if $Y^{l}_i$ is taken between $t_l^s$ and $t_l^e$, we assign it a timestamp $t_l =(t_l^s + t_l^e) / 2$. 
The timestamps are normalized by the total duration of the frame, taking into account the CEP camera's 6~ms readout time, $t_\text{readout}$:
\begin{equation}
t_l' = \frac{t_l}{[\sum_l (t_l^e - t_l^s)] + t_\text{readout}}.
\end{equation}



\noindent
This normalization provides a consistent temporal coordinate $t_l' \in [0,1]$ for every sub-image, facilitating motion interpolation across time.

We use the Real-Time Intermediate Flow Estimation (RIFE) model~\cite{huang2022rife} to estimate bidirectional flows between consecutive sub-images corresponding to the same LED across adjacent frames.
RIFE jointly predicts interpolated flow fields and intermediate frames for arbitrary normalized timesteps $t' \in [0,1]$. 
We specifically use RIFE v4.6~\cite{rifegithub}, which supports arbitrary-timestep warping and high-resolution inference.
Using RIFE, we temporally align all sub-images in a frame to a common reference timestamp. 
We select the lime-colored LED sub-image as the temporal reference because its illumination occurs near the frame's midpoint and its wavelength lies near the spectral center of the LED set.  

Suppose we are aligning the sub-image $Y^{l}_i$ for any LED other than the lime-colored one. 
If LED $l$ is scheduled before the reference ($t'_l < t'_\textit{lime}$), we estimate the interpolated flow fields and warped image between its sub-image in the current frame $Y^l_{i}$ and in the next frame $Y^l_{i+1}$. 
The normalized timestep $t'$ provided to RIFE is determined by the distance between $t'_l$ and $t'_\textit{lime}$,  $t' = t'_\textit{lime} - t'_l$. 
If LED $l$ is scheduled after the reference ($t'_l > t'_\textit{lime}$), we use its sub-image in the previous frame $Y^l_{i-1}$ and in the current frame $Y^l_i$, and the normalized timestep $t'$ provided to RIFE is $t' = (1 - t'_l) + t'_\textit{lime}$. 
We use the warped sub-images produced by RIFE as input to our downstream reconstruction model.

%% file: supplementary/3_reconstruction.tex
\section{Hyperspectral Video Frame Reconstruction}
\label{supp:reconstruction}

To generate the full hyperspectral cube from each coded video frame, we apply hyperspectral reconstruction to the densely coded spectral mosaics. 
This section describes the reconstruction model and its application to the coded video frames.

\subsection{Model Architecture}
\label{supp:architecture}

Our reconstruction network is based on the Holistic Attention Network (HAN)~\cite{niu2020han}, which combines residual learning with hierarchical attention to model inter-channel correlations. The HAN architecture comprises multiple residual groups, each containing convolutional blocks enhanced by layer- and channel-level attention mechanisms. This design effectively captures spatial context and spectral dependencies across the encoded mosaic.

We adapt HAN to our hyperspectral reconstruction setting with three key modifications:

\begin{s_enumerate}
\item We adjust the first and final convolutional layers to handle our input of 12 LED sub-images and output of 33 spectral channels\footnote{Two extrapolated boundary channels are included during training to improve reconstruction near spectral limits but are excluded from evaluation.}.
\item Because our input and output share the same spatial resolution, we remove HAN's original upsampling module.
\item To reduce GPU memory consumption and enable training on a single RTX~A6000 or TITAN~RTX, we reduce the number of residual groups from 20 to 18.
\end{s_enumerate}

The final model contains approximately 57.1M trainable parameters. A detailed \texttt{torchsummary}-style architecture overview is shown in \autoref{fig:han_summary}.

\begin{figure*}[tb]
\centering
{\scriptsize
\begin{minipage}{1\textwidth}
\begin{verbatim}
Layer (type:depth-idx)                         Output Shape           Param #
---------------------------------------------------------------------------------
HANDSA                                         [14, 33, 66, 64]       --
  UnshuffleToSpatiallyPreservingDemosaic: 1-1  [14, 12, 66, 64]       --
    PixelUnshuffle2D: 2-1                      [14, 12, 22, 16]       --
  HAN: 1-2                                     [14, 33, 66, 64]       --
    Sequential: 2-2                            [14, 128, 66, 64]      --
      Conv2d: 3-1                              [14, 128, 66, 64]      13,952
    Sequential: 2-3                            --                     --
      ResidualGroup: 3-2 to 3-11               [14, 128, 66, 64]      54,999,200
      Conv2d: 3-12                             [14, 128, 66, 64]      147,584
    LAM_Module: 2-4                            [14, 1408, 66, 64]     1
      Softmax: 3-13                            [14, 11, 11]           --
    Conv2d: 2-5                                [14, 128, 66, 64]      1,622,144
    CSAM_Module: 2-6                           [14, 128, 66, 64]      1
      Conv3d: 3-14                             [14, 1, 128, 66, 64]   28
      Sigmoid: 3-15                            [14, 1, 128, 66, 64]   --
    Conv2d: 2-7                                [14, 128, 66, 64]      295,040
    Sequential: 2-8                            [14, 33, 66, 64]       --
      Conv2d: 3-16                             [14, 33, 66, 64]       38,049
---------------------------------------------------------------------------------
Total params: 57,115,999 | Trainable: 57,115,999 | Non-trainable: 0
Total mult-adds (T): 3.35
Input size (MB): 9.94 | Fwd/bwd pass (MB): 22726.58 | Params (MB): 228.46
Estimated Total Size (MB): 22964.98
\end{verbatim}
\end{minipage}
}
\caption{A summary of our modified HAN-based reconstruction model.}
\label{fig:han_summary}
\end{figure*}

\subsection{Patch-wise Reconstruction}
\label{supp:patch}

To reduce GPU memory usage and enable efficient training, we reconstruct each frame in smaller overlapping patches. The demosaiced sub-images are partitioned into overlapping $66 \times 64 \times 12$ patches using a sliding window with a stride of $(30,32)$ and no padding. The stride ensures that the top-left corner of each patch aligns with the $3\times4$ spatial–spectral mosaic grid. When the windows do not completely cover the image, additional patches are appended along the right and bottom boundaries.

The HAN network independently processes each patch to reconstruct its corresponding hyperspectral cube. This patch-wise approach allows large full-resolution frames to be reconstructed without exceeding GPU memory limits while maintaining local spectral consistency across overlapping regions.

\subsection{Aggregation of Patch-wise Results}
\label{supp:aggregation}
\begin{figure}[tb]
    \centering
    \includegraphics[width=0.5\linewidth]{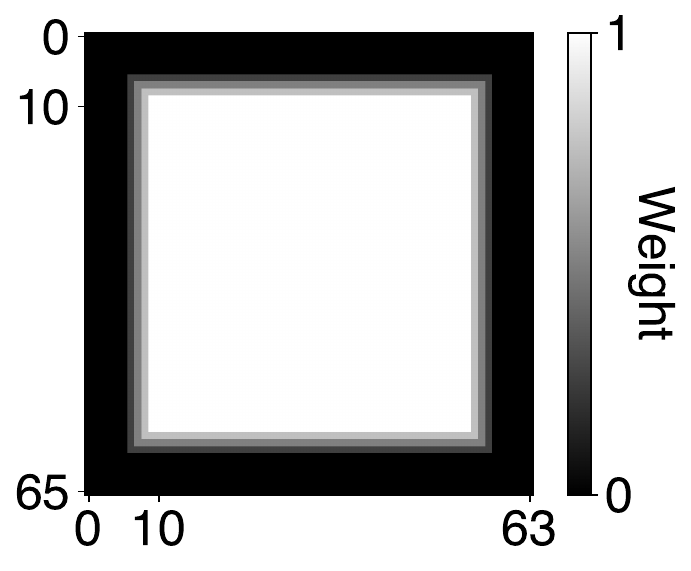}
    \caption{A visualization of the spatial weighting kernel $\mathbf{K}$ used for merging overlapping patch predictions.}
    \label{fig:patch_kernel}
\end{figure}

After reconstructing all patches, we reassemble them into the full-resolution hyperspectral frame using weighted averaging to minimize boundary artifacts. 
Each reconstructed patch is multiplied by a predefined spatial weighting kernel $\mathbf{K} \in [0,1]^{66\times64}$ that assigns higher weights near the center and lower weights near the edges (see \autoref{fig:patch_kernel}). 
This spatial weighting mitigates discontinuities in overlapping regions.

The final full-frame hyperspectral reconstruction is obtained as:
\begin{equation}
\mathbf{R}' = 
\frac{\mathrm{Fold}\!\left(\mathbf{K}_{66 \times 64 \times 33} \odot \mathrm{HAN}(\mathrm{Unfold}(\mathbf{Y}))\right)}
{\mathrm{Fold}(\mathbf{K}_{66 \times 64 \times 33})},
\label{eq:aggregation}
\end{equation}
where the $\odot$ multiplication and the fractional division are applied element-wise. $\mathbf{K}_{66 \times 64 \times 33}$ is the repetition of $\mathbf{K}
 $ in the spectral dimension. $\mathrm{Unfold}:\mathbb{R}^{640\times480} \rightarrow \mathbb{R}^{L\times66\times64}$ and $\mathrm{Fold}:\mathbb{R}^{L\times66\times64\times33} \rightarrow \mathbb{R}^{640\times480\times33} $ are inverse-like sliding-window operations used to extract and reassemble the $L$ overlapping patches, respectively. 
$\mathrm{HAN}: \mathbb{R}^{66 \times 64} \rightarrow \mathbb{R}^{66 \times 64 \times 33}$ denotes the learned Holistic Attention Network used for patch-wise hyperspectral reconstruction.

%% file: supplementary/4_additional_results_simulations.tex
\section{Additional Results: Simulations}
\label{supp:results-simulations}

\label{supp:training-details}
\subsection{Additional Details}
During training, we extract random patches from each simulated measurement. To ensure correct alignment between the illumination schedule and the pixel-wise coded-exposure pattern, the top-left corner of every patch is constrained to coincide with the top-left corner of a $3 \times 4$ mosaic tile (see \autoref{fig:mosaic}). This guarantees that every extracted patch contains an integer number of complete mosaic repetitions and avoids boundary inconsistencies during reconstruction.
Each batch is formed by sampling patches independently—potentially from different hyperspectral images—to improve spectral diversity and reduce overfitting. When noise robustness is evaluated, we add independent Gaussian noise with standard deviation $\sigma \in {0,5,10,15,20}\%$ of the maximum signal to each simulated patch before feeding it to the network.
All corresponding hyperspectral ground-truth patches are extracted with identical spatial coordinates and normalized consistently. This pipeline ensures that the model learns reconstruction mappings that are faithful to the physical sensing process used in the real hardware.

\subsection{Comparisons with Baselines}
\label{supp:baselines}
\begin{figure*}[p]
    \centering
    \includegraphics[width=0.98\textwidth]{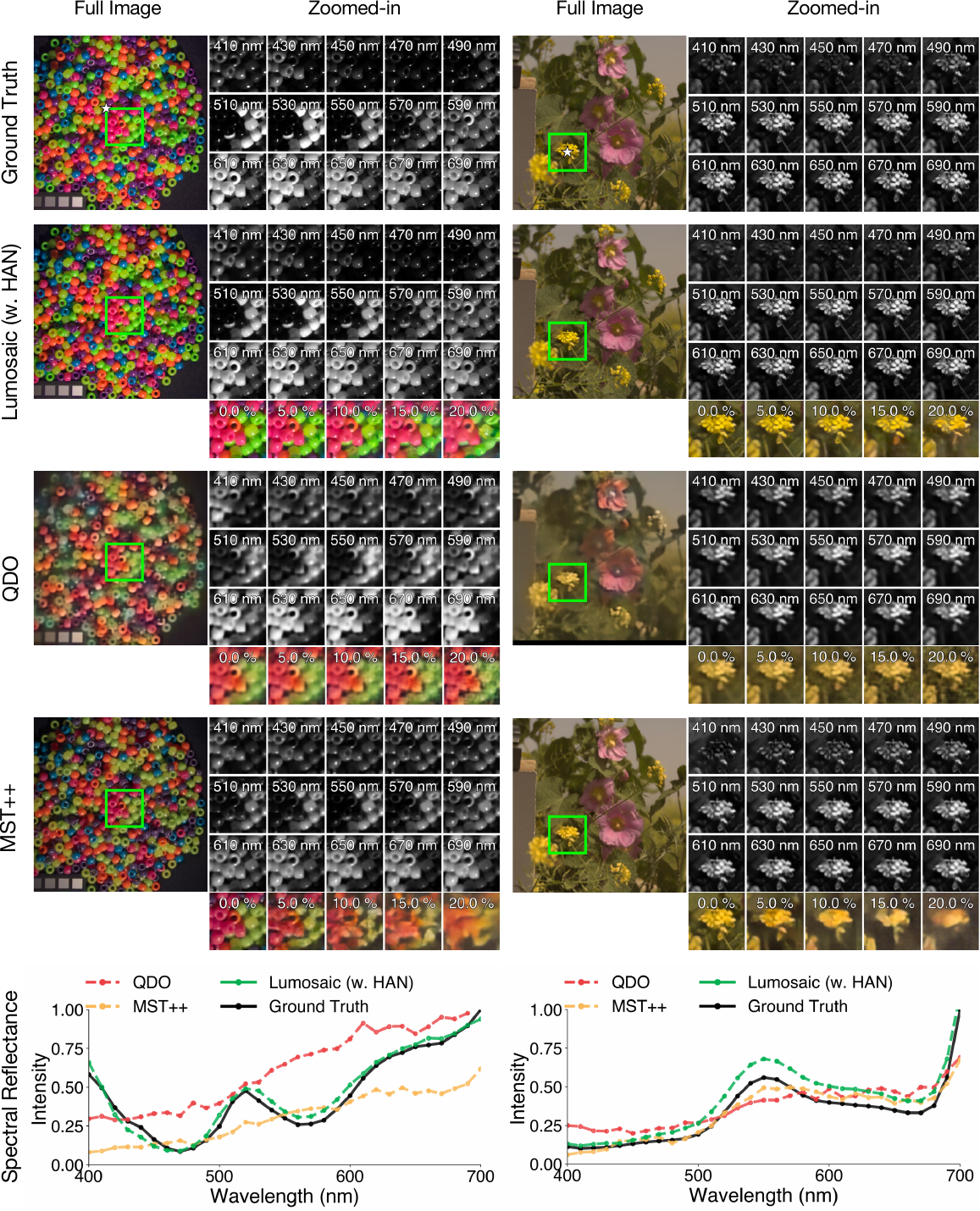}
    \caption{
    \textbf{Qualitative comparisons on public datasets.}
    Representative reconstructions from the test set comparing QDO~\cite{li2022qdo}, MST++~\cite{cai2022mstplusplus}, and \projectname.
    Each column shows the reconstructed sRGB image, selected hyperspectral channels, 
    and zoomed-in sRGB crops under varying Gaussian noise levels (annotated with the 
    corresponding standard deviation). 
    We also present spectral validation plots for two specific locations, marked with white stars in the first-row sRGB images, displayed at the bottom of the figure.
    }
    \label{fig:simulations-qual-supp}
\end{figure*}

\autoref{fig:simulations-qual-supp} presents qualitative comparisons of \projectname with a HAN backbone on test scenes from the CAVE~\cite{yasuma2008cave}, KAUST~\cite{li2021kaust}, and ARAD~\cite{arad2022recovery} datasets.
Reconstructed outputs from QDO~\cite{li2022qdo} and MST++~\cite{cai2022mstplusplus} are provided as state-of-the-art comparisons.
Across all the scenes presented, \projectname recovers finer spectral details and spatial textures while avoiding color bleeding commonly observed in the baselines.

\FloatBarrier

\subsection{High-Frequency Spectral Recovery Test}
\label{supp:spectral-resolution}

\begin{figure*}[p]
    \centering
    \includegraphics[width=0.9\textwidth]{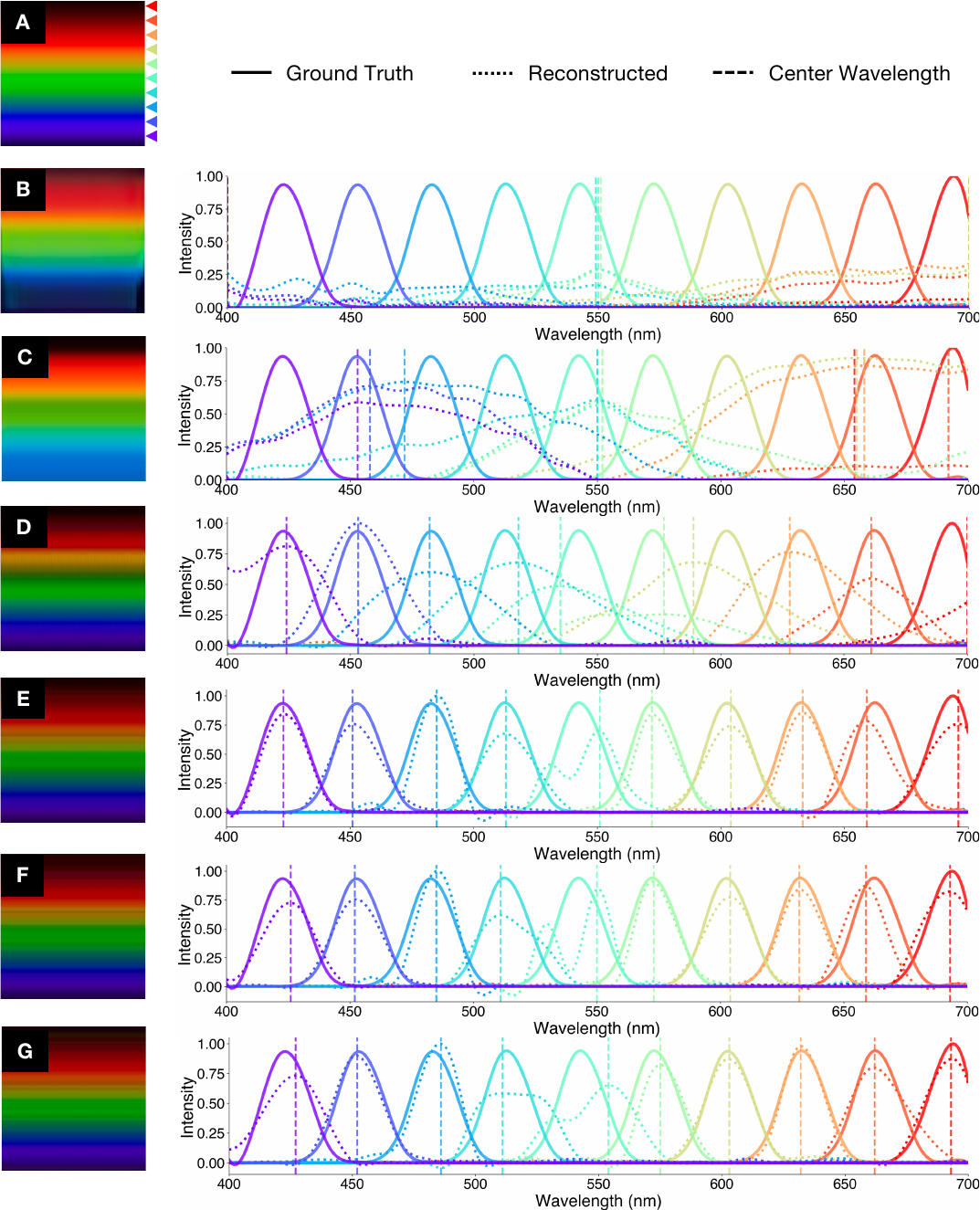}
    \caption{
    \textbf{Spectral recovery test on a high-frequency rainbow scene.}
    (\textbf{Left}) sRGB visualization of the rainbow scene: (A) ground truth and reconstructions from (B) QDO~\cite{li2022qdo}, (C) MST++~\cite{cai2022mstplusplus}, and (D) \projectname.
    Rows (D-F) illustrate 
    \projectname results under increasing proportions of high-frequency synthetic spectra in the training set (0\%, 33\%, 66\%, 100\%). 
    (\textbf{Right}) Reconstructed spectral reflectance, interpolated at 1-nm intervals, and center wavelength positions at selected locations on the rainbow (marked by triangles), compared to ground truth, demonstrating \projectname's improved ability to recover sharp spectral transitions. }
    \label{fig:spectral-resolution-supp}
\end{figure*}

We assess \projectname's ability to recover high-frequency spectral features using a synthetic but deliberately challenging rainbow scene (\autoref{fig:spectral-resolution-supp}).
The scene is created by sweeping Gaussian-shaped spectral reflectance profiles with a FWHM of 20\,nm\footnote{We choose the narrowest FWHM supported by the LEDs used in our implementation.} from the bottom to the top of a $512 \times 512$ image.
The gradient spans the 400–700\,nm range, so each row's central wavelength changes by (700-400\,nm) / 512~px = 0.59\,nm/px.
Thus, adjacent rows differ only slightly in their peak wavelength, producing a high-frequency spectral gradient that stresses the model’s ability to resolve fine spectral variations.
This setup offers a controlled benchmark containing sharp spectral transitions that are rarely observed in natural scenes. As a result, it provides a practical out-of-distribution test of spectral resolving capability, revealing failure modes in conventional snapshot systems that rely heavily on natural-scene priors and tend to oversmooth high-frequency spectra.

As shown in \autoref{fig:spectral-resolution-supp}(A-D), the baseline QDO~\cite{li2022qdo} and MST++~\cite{cai2022mstplusplus} methods exhibit noticeable spectral smoothing and blending artifacts, failing to reproduce abrupt transitions between neighboring wavelength bands.
In contrast, \projectname reconstructs these features while maintaining acceptable spatial fidelity and substantially better spectral localization across most of the gradient. 
This improvement stems from its deterministic mosaic-based sensing strategy, which captures densely sampled, non-multiplexed spectral measurements rather than entangling multiple wavelengths within a single coded exposure.
By reducing reliance on strong learned priors and providing cleaner spectral cues directly in the measurement domain, \projectname\ achieves more accurate and reliable reconstruction of fine spectral structure.

We further analyze the influence of training data composition by introducing synthetic spectra that better represent narrow-band spectral features. 
The synthetic training data include two types of spectral vectors: 
\begin{enumerate}
\item \textbf{Single-peak profiles:} Gaussian functions are generated for center wavelengths from 400–700\,nm with FWHM values of \{10, 20, 30, 40, 50\}\,nm. 
Each profile is sampled at 1-nm intervals, integrated into 10-nm bins using trapezoidal integration, and normalized to a maximum value of~1. 
\item \textbf{Double-peak profiles:} Two Gaussian profiles are combined with center separations of \{10, 20, 30, 40, 60, 80\}\,nm, ensuring both peaks remain between 400 and 700\,nm; the same integration and normalization procedure is applied. 
\end{enumerate}
For training purposes, we use these synthetic spectra to create uniform, textureless hyperspectral images by repeating each spectral vector spatially. 

We compare versions of \projectname trained on 0\%, 33\%, 66\%, and 100\% synthetic spectra mixed with to natural scenes from our public dataset corpus, and and re-evaluate them on the rainbow scene. 
As shown in \autoref{fig:spectral-resolution-supp}(D-G), moderate synthetic augmentation (33\%) provides the best trade-off between generalization and spectral precision, producing sharper and well-localized spectral peaks that closely align with the ground-truth spectra. 
Models trained exclusively on natural scenes exhibit lower sensitivity to subtle spectral variations, while heavier synthetic mixing (66\% and 100\%) offers limited additional resolving power and can slightly bias the reconstructions toward narrowband profiles.

\FloatBarrier

%% file: supplementary/5_additional_results_real_world.tex
\section{Additional Results: Real World}
\label{supp:results-realworld}

\subsection{Static Scenes}
\label{supp:static-scenes}

\begin{figure*}[!t]
    \centering
  \includegraphics[width=1\textwidth]{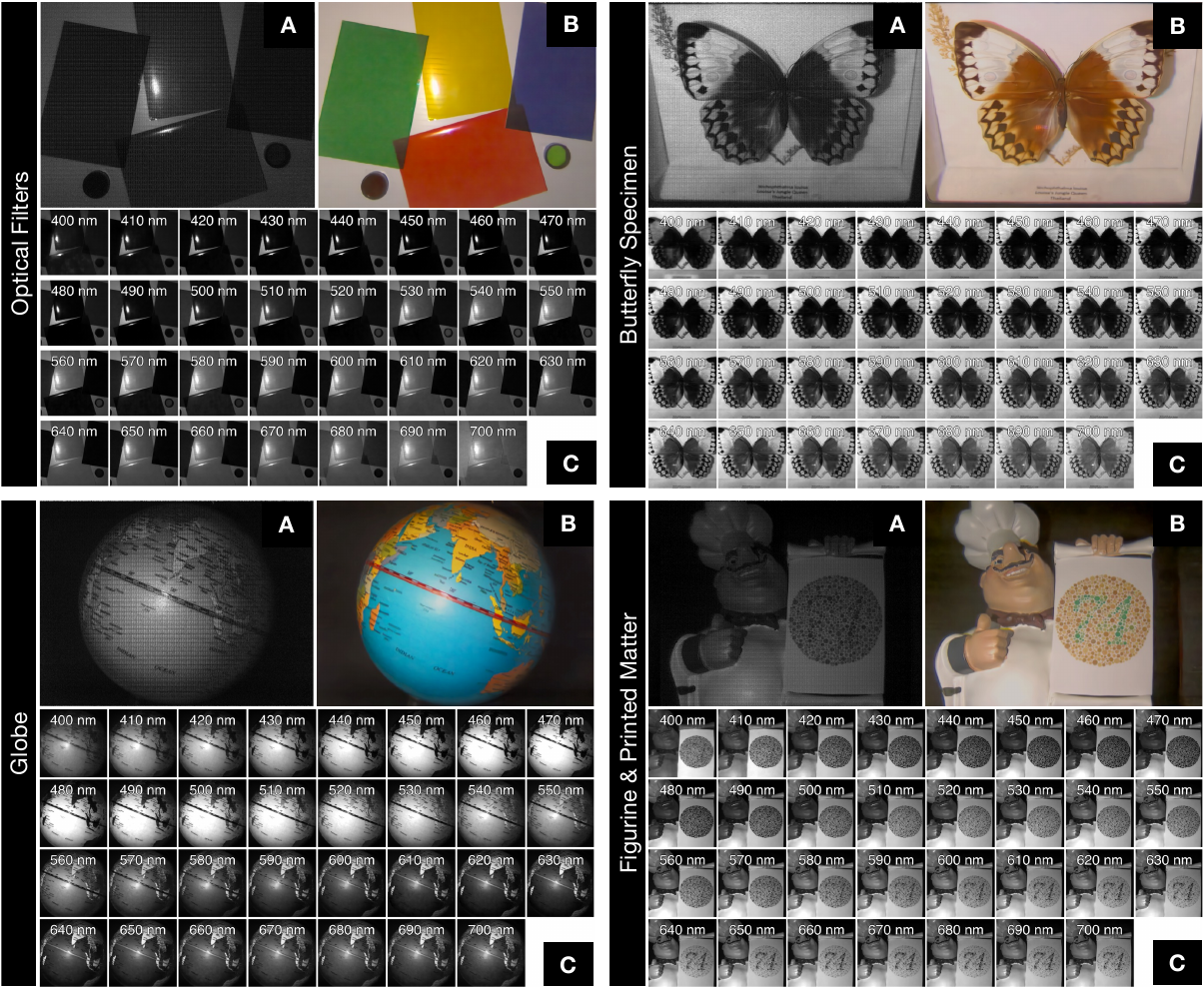}
  \caption{
    Additional \projectname reconstruction results on diverse static scenes: optical filters, a globe, a butterfly specimen, and a figurine with printed matter. Each scene includes \textbf{(A)} the input coded image, \textbf{(B)} the reconstructed hyperspectral image rendered as sRGB, and \textbf{(C)} the 31-channel hyperspectral composites rendered in grayscale for visualization.}
  \label{fig:static-scenes-suppl}
\end{figure*}

\autoref{fig:static-scenes-suppl} further demonstrates \projectname's hyperspectral reconstruction capabilities for static scenes.
As shown, \projectname correctly reconstructs fine-grained spatial details in each scene: the names of major countries and bodies of water on the globe, the box label and thin branches surrounding the butterfly, and the teeth of the figurine.
The spectral information is also accurate, as clearly illustrated by the distinct colors of the countries on the globe.

\subsection{Metamerism Analysis}
\label{supp:metamerism}

\begin{figure*}[htbp!]
    \centering
    \includegraphics[width=0.7\textwidth]{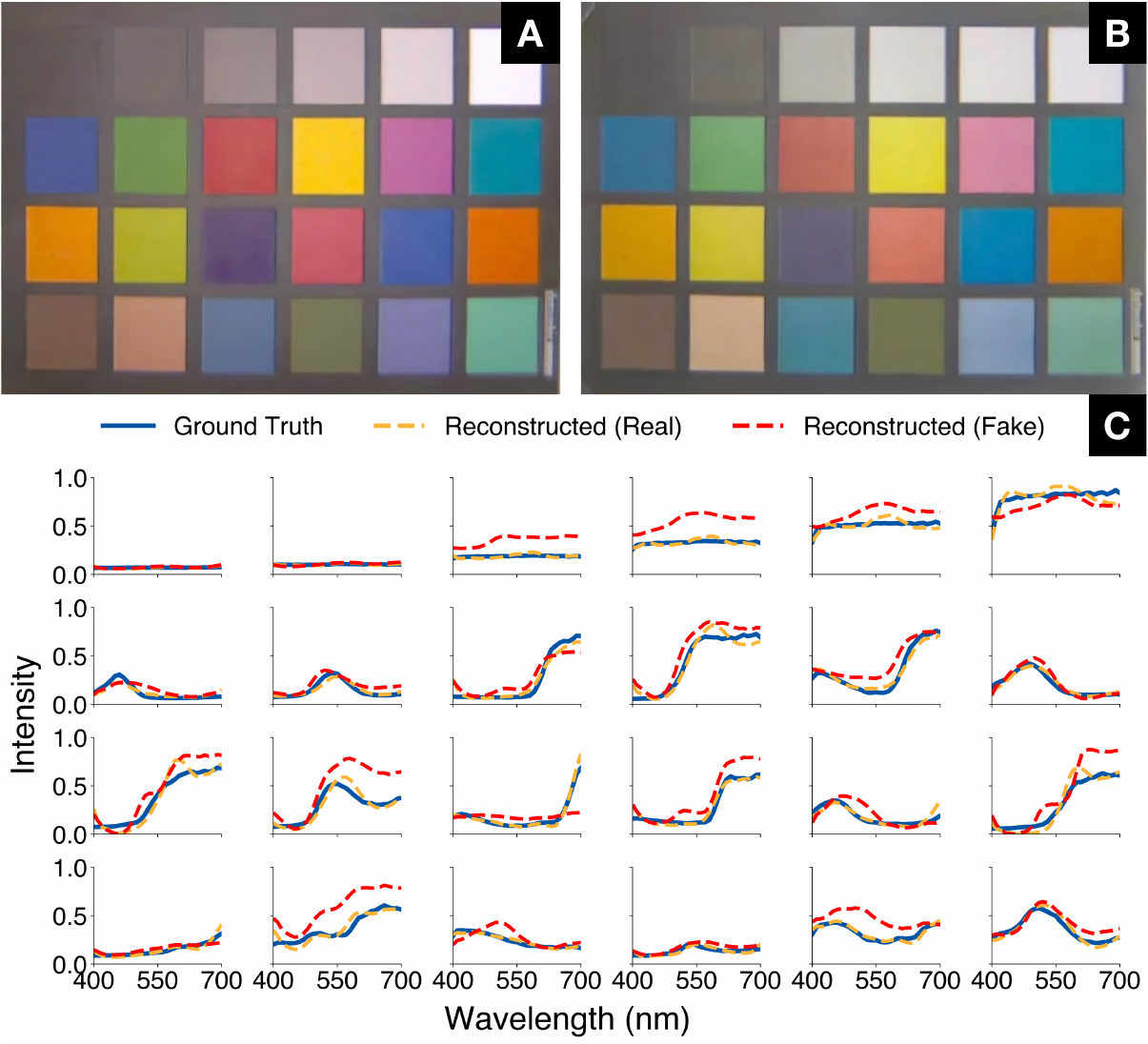}
    \caption{
    \textbf{Metamerism analysis with \projectname using genuine and printed ColorChecker targets.}
    (\textbf{A and B}) RGB visualization of the genuine ColorChecker and a printed photocopy under identical illumination.
    (\textbf{C}) Spectral curves illustrating the differences between genuine and printed patches. 
    }
    \label{fig:metamerism_results}
\end{figure*}

\autoref{fig:metamerism_results} showcases \projectname's ability to resolve metameric ambiguities. 
Both targets appear visually identical under sRGB rendering, yet their spectral profiles differ notably across multiple patches.
The genuine ColorChecker exhibits smooth, well-defined reflectance spectra characteristic of pigmented surfaces, whereas the printed copy shows irregular spectral peaks due to ink absorption and printer gamut limitations. 

\subsection{Dynamic Scenes}
\label{supp:dynamic-scenes}

\begin{figure*}[p]
    \centering
  \includegraphics[width=0.75\textwidth]{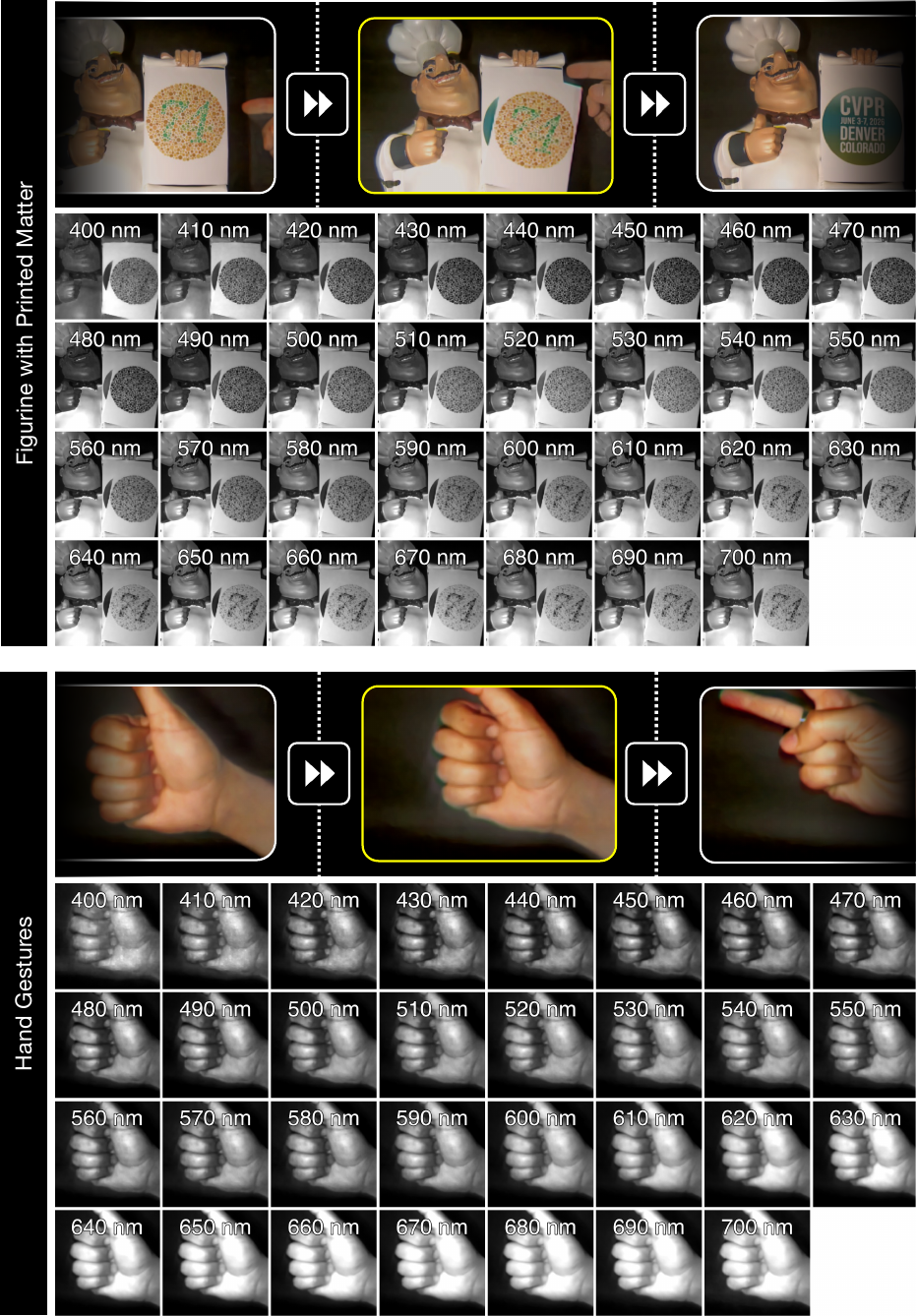}
  \caption{
    Additional hyperspectral reconstruction results with \projectname on scenes exhibiting diverse motion characteristics. 
    \textbf{(top)} A figurine with printed elements, including Ishihara patterns and text, undergoing both rotational and falling motion relative to the camera. 
    \textbf{(bottom)} Dynamic hand gestures with non-rigid motion. 
    The top row for each scene shows rendered sRGB views from non-consecutive frames of the reconstructed hyperspectral video, while the bottom row shows the corresponding full 31-channel hyperspectral images for the frames outlined in yellow.
}
  \label{fig:dynamic-scenes-suppl1}
\end{figure*}

\begin{figure*}[p]
    \centering
    \includegraphics[width=0.75\textwidth]{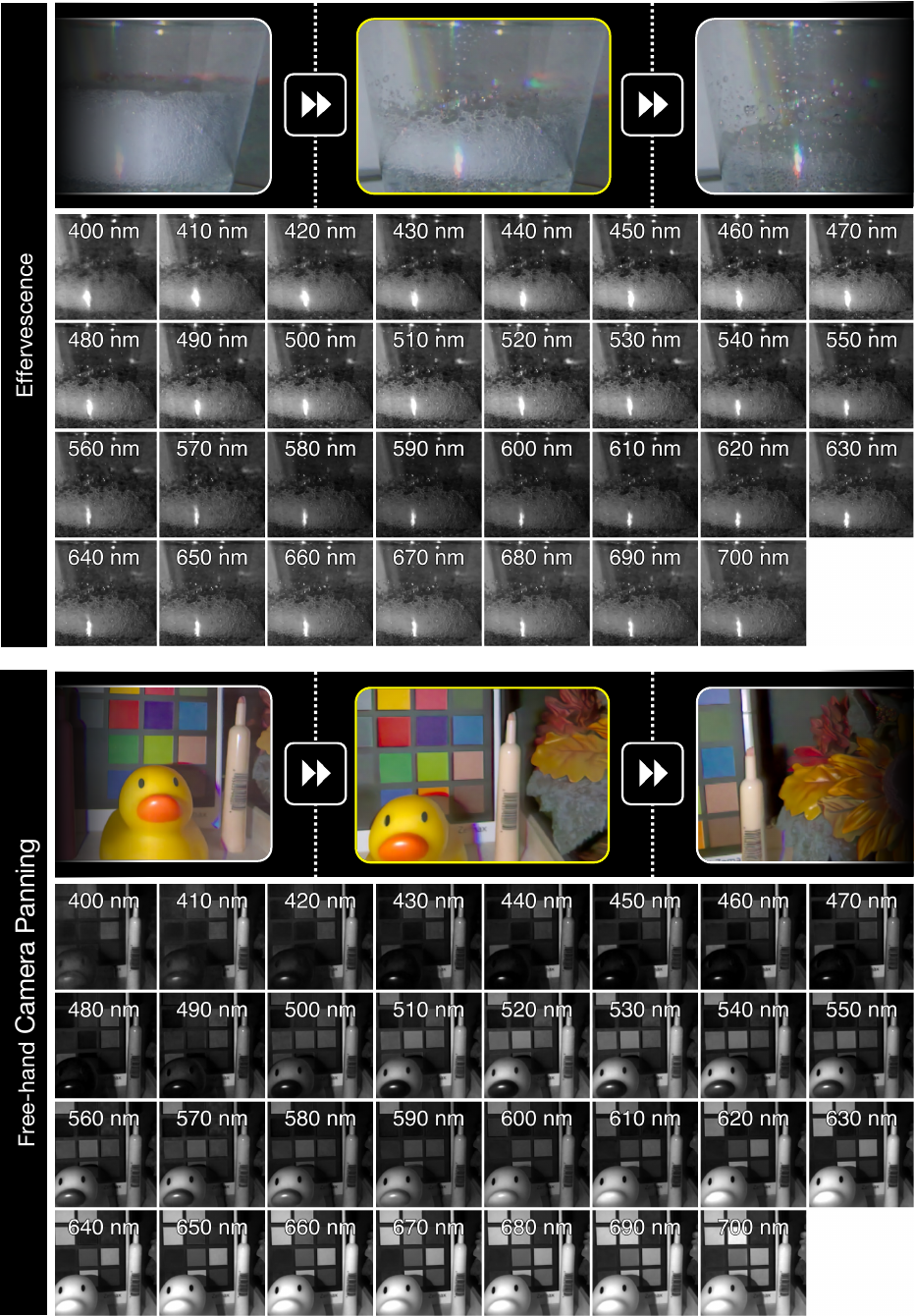}
    \caption{
        Additional hyperspectral reconstruction results with \projectname on scenes exhibiting diverse motion characteristics. 
        \textbf{(top)} A transparent plastic cup filled with carbonated tonic water, capturing high-frequency motion due to effervescence. 
        \textbf{(bottom)} A complex indoor scene with colorful objects captured while freely panning the imaging system to introduce continuous viewpoint and illumination changes. 
        The top row for each scene shows rendered sRGB views sampled from non-consecutive frames of the reconstructed hyperspectral video, while the bottom row shows the corresponding full 31-channel hyperspectral images for the frames outlined in red.
    }
    \label{fig:dynamic-scenes-suppl2}
\end{figure*}

\autoref{fig:dynamic-scenes-suppl1} and \autoref{fig:dynamic-scenes-suppl2} further demonstrate \projectname's capabilities as a hyperspectral video reconstruction system for dynamic scenes.
In the scene with the rotating figurine (\autoref{fig:dynamic-scenes-suppl1}, top), all printed elements, including the Ishihara pattern and the conference advertisement is properly resolved across the entire video, despite continuous rotational motion.
In the hand gesture scene (\autoref{fig:dynamic-scenes-suppl1}, bottom), there is some slight ghosting when the hand motion is fastest, but the content is nonetheless coherent. 
The scene with effervescent tonic water (\autoref{fig:dynamic-scenes-suppl2}, top) is particularly well-suited for evaluating hyperspectral video, as tonic water exhibits strong fluorescence for an otherwise transparent liquid. 
Despite the liquid's transparency, the system is able to render the bubbles with high spatial resolution.
Finally, we demonstrate robustness to camera motion through a free-hand panning sequence (\autoref{fig:dynamic-scenes-suppl2}, bottom), in which the system maintains stable hyperspectral reconstruction despite rapid viewpoint changes.

\subsection{Ablation: Effect of Temporal Alignment}
\label{supp:ablation-temporal}

\begin{figure*}[htbp]
    \centering
    \includegraphics[width=0.7\textwidth]{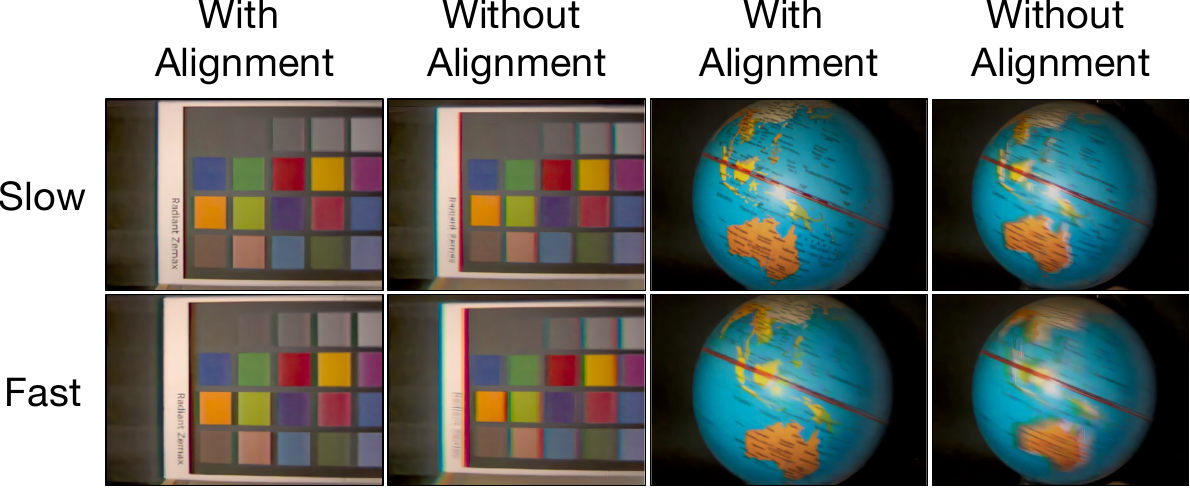}
    \caption{
        The effect of temporal alignment on \projectname's reconstruction quality in dynamic scenes. 
        We compare results on two motion regimes—\textbf{(top)} slow and \textbf{(bottom)} fast—showing hyperspectral reconstructions rendered in sRGB with and without temporal alignment.
    }\label{fig:temporal_alignment}
    \vspace{-0.2in}
\end{figure*}

We assess the importance of our temporal alignment module by visualizing \projectname reconstructions with and without alignment at varying motion speeds. 
As shown in \autoref{fig:temporal_alignment}, fast-moving scenes exhibit ghosting and spectral blending artifacts when temporal alignment is not in use. 
These artifacts become increasingly pronounced at higher motion speeds, especially around edges or in areas with large brightness gradients. 
Applying temporal alignment produces sharper and more spectrally consistent outputs, confirming its crucial role in maintaining fidelity under motion.
This is only made possible by the temporally coded illumination–exposure scheme underlying \projectname. 
Because each sub-frame captures a distinct wavelength band at a known temporal offset, our system inherently encodes both spectral and motion information within the same exposure sequence. 
The alignment module leverages this temporal structure to compensate for inter-frame motion, effectively restoring spatial–spectral coherence in dynamic scenes.

%% file: ref.bib
@String(CVPR= {IEEE Conf. Comput. Vis. Pattern Recog.})

@String(ECCV= {Eur. Conf. Comput. Vis.})

@String(TOG= {ACM Trans. Graph.})

@String(CVPR  = {CVPR})

@String(ECCV  = {ECCV})

@String(TOG   = {ACM TOG})

@inproceedings{zhi2019multispectral,
  title={Multispectral imaging for fine-grained recognition of powders on complex backgrounds},
  author={Zhi, Tiancheng and Pires, Bernardo R and Hebert, Martial and Narasimhan, Srinivasa G},
  booktitle={Proceedings of the IEEE/CVF Conference on Computer Vision and Pattern Recognition},
  pages={8699--8708},
  year={2019}
}

@article{lu2014medical,
  title={Medical hyperspectral imaging: a review},
  author={Lu, Guolan and Fei, Baowei},
  journal={Journal of biomedical optics},
  volume={19},
  number={1},
  pages={010901--010901},
  year={2014},
  publisher={Society of Photo-Optical Instrumentation Engineers}
}

@article{legendre2016practical,
  title={Practical multispectral lighting reproduction},
  author={LeGendre, Chloe and Yu, Xueming and Liu, Dai and Busch, Jay and Jones, Andrew and Pattanaik, Sumanta and Debevec, Paul},
  journal={ACM Transactions on Graphics (TOG)},
  volume={35},
  number={4},
  pages={1--11},
  year={2016},
  publisher={ACM New York, NY, USA}
}

@article{arce2013compressive,
  title={Compressive coded aperture spectral imaging: An introduction},
  author={Arce, Gonzalo R and Brady, David J and Carin, Lawrence and Arguello, Henry and Kittle, David S},
  journal={IEEE Signal Processing Magazine},
  volume={31},
  number={1},
  pages={105--115},
  year={2013},
  publisher={IEEE}
}

@inproceedings{baek2021single,
  title={Single-shot hyperspectral-depth imaging with learned diffractive optics},
  author={Baek, Seung-Hwan and Ikoma, Hayato and Jeon, Daniel S and Li, Yuqi and Heidrich, Wolfgang and Wetzstein, Gordon and Kim, Min H},
  booktitle={Proceedings of the IEEE/CVF International Conference on Computer Vision},
  pages={2651--2660},
  year={2021}
}

@inproceedings{geelen2014compact,
  title={A compact snapshot multispectral imager with a monolithically integrated per-pixel filter mosaic},
  author={Geelen, Bert and Tack, Nicolaas and Lambrechts, Andy},
  booktitle={Advanced fabrication technologies for micro/nano optics and photonics VII},
  volume={8974},
  pages={80--87},
  year={2014},
  organization={SPIE}
}

@inproceedings{park2007multispectral,
  title={Multispectral imaging using multiplexed illumination},
  author={Park, Jong-Il and Lee, Moon-Hyun and Grossberg, Michael D and Nayar, Shree K},
  booktitle={2007 IEEE 11th International Conference on Computer Vision},
  pages={1--8},
  year={2007},
  organization={IEEE}
}

@article{shin2024dense,
  title={Dense Dispersed Structured Light for Hyperspectral 3D Imaging of Dynamic Scenes},
  author={Shin, Suhyun and Yoon, Seungwoo and Maeda, Ryota and Baek, Seung-Hwan},
  journal={arXiv preprint arXiv:2412.01140},
  year={2024}
}

@article{bub2010temporal,
  title={Temporal pixel multiplexing for simultaneous high-speed, high-resolution imaging},
  author={Bub, Gil and Tecza, Matthias and Helmes, Michiel and Lee, Peter and Kohl, Peter},
  journal={Nature methods},
  volume={7},
  number={3},
  pages={209--211},
  year={2010},
  publisher={Nature Publishing Group US New York}
}

@article{kang2022indirect,
  title={An indirect time-of-flight sensor with tetra-pixel architecture calibrating tap mismatch in a single frame},
  author={Kang, Jubin and Park, Yongjae and Hwang, Jung-Hye and Hong, Kieop and Son, Insang and Chun, Jung-Hoon and Choi, Jaehyuk and Kim, Seong-Jin},
  journal={IEEE Solid-State Circuits Letters},
  volume={5},
  pages={284--287},
  year={2022},
  publisher={IEEE}
}

@inproceedings{wei2018coded,
  title={Coded two-bucket cameras for computer vision},
  author={Wei, Mian and Sarhangnejad, Navid and Xia, Zhengfan and Gusev, Nikita and Katic, Nikola and Genov, Roman and Kutulakos, Kiriakos N},
  booktitle={Proceedings of the European Conference on Computer Vision (ECCV)},
  pages={54--71},
  year={2018}
}

@article{verma2024chromaflash,
  title={ChromaFlash: Snapshot Hyperspectral Imaging Using Rolling Shutter Cameras},
  author={Verma, Dhruv and Ruffolo, Ian and Lindell, David B and Kutulakos, Kiriakos N and Mariakakis, Alex},
  journal={Proceedings of the ACM on Interactive, Mobile, Wearable and Ubiquitous Technologies},
  volume={8},
  number={3},
  pages={1--31},
  year={2024},
  publisher={ACM New York, NY, USA}
}

@inproceedings{goel2015hypercam,
  title={HyperCam: hyperspectral imaging for ubiquitous computing applications},
  author={Goel, Mayank and Whitmire, Eric and Mariakakis, Alex and Saponas, T Scott and Joshi, Neel and Morris, Dan and Guenter, Brian and Gavriliu, Marcel and Borriello, Gaetano and Patel, Shwetak N},
  booktitle={Proceedings of the 2015 ACM International Joint Conference on Pervasive and Ubiquitous Computing},
  pages={145--156},
  year={2015}
}

@inproceedings{li2019pro,
  title={Pro-cam ssfm: Projector-camera system for structure and spectral reflectance from motion},
  author={Li, Chunyu and Monno, Yusuke and Hidaka, Hironori and Okutomi, Masatoshi},
  booktitle={Proceedings of the IEEE/CVF International Conference on Computer Vision},
  pages={2414--2423},
  year={2019}
}

@inproceedings{makarenko2022real,
  title={Real-time hyperspectral imaging in hardware via trained metasurface encoders},
  author={Makarenko, Maksim and Burguete-Lopez, Arturo and Wang, Qizhou and Getman, Fedor and Giancola, Silvio and Ghanem, Bernard and Fratalocchi, Andrea},
  booktitle={Proceedings of the IEEE/CVF Conference on Computer Vision and Pattern Recognition},
  pages={12692--12702},
  year={2022}
}

@article{baek2017compact,
  title={Compact single-shot hyperspectral imaging using a prism},
  author={Baek, Seung-Hwan and Kim, Incheol and Gutierrez, Diego and Kim, Min H},
  journal={ACM Transactions on Graphics (TOG)},
  volume={36},
  number={6},
  pages={1--12},
  year={2017},
  publisher={ACM New York, NY, USA}
}

@inproceedings{zhao2019hyperspectral,
  title={Hyperspectral imaging with random printed mask},
  author={Zhao, Yuanyuan and Guo, Hui and Ma, Zhan and Cao, Xun and Yue, Tao and Hu, Xuemei},
  booktitle={Proceedings of the IEEE/CVF Conference on Computer Vision and Pattern Recognition},
  pages={10149--10157},
  year={2019}
}

@article{shi2024learned,
  title={Learned Multi-aperture Color-coded Optics for Snapshot Hyperspectral Imaging},
  author={Shi, Zheng and Dun, Xiong and Wei, Haoyu and Dong, Siyu and Wang, Zhanshan and Cheng, Xinbin and Heide, Felix and Peng, Yifan},
  journal={ACM Transactions on Graphics (TOG)},
  volume={43},
  number={6},
  pages={1--11},
  year={2024},
  publisher={ACM New York, NY, USA}
}

@inproceedings{shin2024dispersed,
  title={Dispersed structured light for hyperspectral 3d imaging},
  author={Shin, Suhyun and Choi, Seokjun and Heide, Felix and Baek, Seung-Hwan},
  booktitle={Proceedings of the IEEE/CVF Conference on Computer Vision and Pattern Recognition},
  pages={24997--25006},
  year={2024}
}

@inproceedings{li2022deep,
  title={Deep hyperspectral-depth reconstruction using single color-dot projection},
  author={Li, Chunyu and Monno, Yusuke and Okutomi, Masatoshi},
  booktitle={Proceedings of the IEEE/CVF Conference on Computer Vision and Pattern Recognition},
  pages={19770--19779},
  year={2022}
}

@article{gulve202339,
  title={39 000-subexposures/s Dual-ADC CMOS image sensor with dual-tap coded-exposure pixels for single-shot HDR and 3-D computational imaging},
  author={Gulve, Rahul and Sarhangnejad, Navid and Dutta, Gairik and Sakr, Motasem and Nguyen, Don and Rangel, Roberto and Chen, Wenzheng and Xia, Zhengfan and Wei, Mian and Gusev, Nikita and others},
  journal={IEEE Journal of Solid-State Circuits},
  volume={58},
  number={11},
  pages={3150--3163},
  year={2023},
  publisher={IEEE}
}

@inproceedings{huang2022rife,
author = {Huang, Zhewei and Zhang, Tianyuan and Heng, Wen and Shi, Boxin and Zhou, Shuchang},
title = {Real-Time Intermediate Flow Estimation for Video Frame Interpolation},
year = {2022},
isbn = {978-3-031-19780-2},
publisher = {Springer-Verlag},
address = {Berlin, Heidelberg},
url = {https://doi.org/10.1007/978-3-031-19781-9_36},
doi = {10.1007/978-3-031-19781-9_36},
booktitle = {Computer Vision – ECCV 2022: 17th European Conference, Tel Aviv, Israel, October 23–27, 2022, Proceedings, Part XIV},
pages = {624–642},
numpages = {19},
location = {Tel Aviv, Israel}
}

@misc{rifegithub,
  author = {Huang, Zhewei and Zhang, Tianyuan and Heng, Wen and Shi, Boxin and Zhou, Shuchang},
  title = {Real-Time Intermediate Flow Estimation for Video Frame Interpolation},
  year = {2022},
  publisher = {GitHub},
  journal = {GitHub repository},
  howpublished = {\url{https://github.com/hzwer/ECCV2022-RIFE}},
  commit = {faf78cd5a139d454b3724a0e67f5bb4a1943c1ac}
}

@inproceedings{niu2020han,
author = {Niu, Ben and Wen, Weilei and Ren, Wenqi and Zhang, Xiangde and Yang, Lianping and Wang, Shuzhen and Zhang, Kaihao and Cao, Xiaochun and Shen, Haifeng},
title = {Single Image Super-Resolution via a Holistic Attention Network},
year = {2020},
isbn = {978-3-030-58609-6},
publisher = {Springer-Verlag},
address = {Berlin, Heidelberg},
url = {https://doi.org/10.1007/978-3-030-58610-2_12},
doi = {10.1007/978-3-030-58610-2_12},
booktitle = {Computer Vision – ECCV 2020: 16th European Conference, Glasgow, UK, August 23–28, 2020, Proceedings, Part XII},
pages = {191–207},
numpages = {17},
location = {Glasgow, United Kingdom}
}

@misc{li2021kaust,
  title     = "Dataset for multispectral illumination estimation using deep
               unrolling network",
  author    = "Li, Yuqi and Fu, Qiang and Heidrich, Wolfgang",
  publisher = "KAUST Research Repository",
  year      =  2021
}

@article{yasuma2008cave,
  title={Generalized assorted pixel camera: postcapture control of resolution, dynamic range, and spectrum},
  author={Yasuma, Fumihito and Mitsunaga, Tomoo and Iso, Daisuke and Nayar, Shree K},
  journal={IEEE transactions on image processing},
  volume={19},
  number={9},
  pages={2241--2253},
  year={2010},
  publisher={IEEE}
}

@InProceedings{arad2022recovery,
    author    = {Arad, Boaz and Timofte, Radu and Yahel, Rony and Morag, Nimrod and Bernat, Amir and Cai, Yuanhao and Lin, Jing and Lin, Zudi and Wang, Haoqian and Zhang, Yulun and Pfister, Hanspeter and Van Gool, Luc and Liu, Shuai and Li, Yongqiang and Feng, Chaoyu and Lei, Lei and Li, Jiaojiao and Du, Songcheng and Wu, Chaoxiong and Leng, Yihong and Song, Rui and Zhang, Mingwei and Song, Chongxing and Zhao, Shuyi and Lang, Zhiqiang and Wei, Wei and Zhang, Lei and Dian, Renwei and Shan, Tianci and Guo, Anjing and Feng, Chengguo and Liu, Jinyang and Agarla, Mirko and Bianco, Simone and Buzzelli, Marco and Celona, Luigi and Schettini, Raimondo and He, Jiang and Xiao, Yi and Xiao, Jiajun and Yuan, Qiangqiang and Li, Jie and Zhang, Liangpei and Kwon, Taesung and Ryu, Dohoon and Bae, Hyokyoung and Yang, Hao-Hsiang and Chang, Hua-En and Huang, Zhi-Kai and Chen, Wei-Ting and Kuo, Sy-Yen and Chen, Junyu and Li, Haiwei and Liu, Song and Sabarinathan and Uma, K and Bama, B Sathya and Roomi, S. Mohamed Mansoor},
    title     = {NTIRE 2022 Spectral Recovery Challenge and Data Set},
    booktitle = {Proceedings of the IEEE/CVF Conference on Computer Vision and Pattern Recognition (CVPR) Workshops},
    month     = {June},
    year      = {2022},
    pages     = {863-881}
}

@InProceedings{cai2022mstplusplus,
    author    = {Cai, Yuanhao and Lin, Jing and Lin, Zudi and Wang, Haoqian and Zhang, Yulun and Pfister, Hanspeter and Timofte, Radu and Van Gool, Luc},
    title     = {MST++: Multi-Stage Spectral-Wise Transformer for Efficient Spectral Reconstruction},
    booktitle = {Proceedings of the IEEE/CVF Conference on Computer Vision and Pattern Recognition (CVPR) Workshops},
    month     = {June},
    year      = {2022},
    pages     = {745-755}
}

@INPROCEEDINGS{li2022qdo,
  author={Li, Lingen and Wang, Lizhi and Song, Weitao and Zhang, Lei and Xiong, Zhiwei and Huang, Hua},
  booktitle={2022 IEEE/CVF Conference on Computer Vision and Pattern Recognition (CVPR)}, 
  title={Quantization-aware Deep Optics for Diffractive Snapshot Hyperspectral Imaging}, 
  year={2022},
  volume={},
  number={},
  pages={19748-19757},
  doi={10.1109/CVPR52688.2022.01916}}

@article{feng2021mosaic,
  title={Mosaic convolution-attention network for demosaicing multispectral filter array images},
  author={Feng, Kai and Zhao, Yongqiang and Chan, Jonathan Cheung-Wai and Kong, Seong G and Zhang, Xun and Wang, Binglu},
  journal={IEEE Transactions on Computational Imaging},
  volume={7},
  pages={864--878},
  year={2021},
  publisher={IEEE}
}

@article{bian2024broadband,
  title={A broadband hyperspectral image sensor with high spatio-temporal resolution},
  author={Bian, Liheng and Wang, Zhen and Zhang, Yuzhe and Li, Lianjie and Zhang, Yinuo and Yang, Chen and Fang, Wen and Zhao, Jiajun and Zhu, Chunli and Meng, Qinghao and others},
  journal={Nature},
  volume={635},
  number={8037},
  pages={73--81},
  year={2024},
  publisher={Nature Publishing Group UK London}
}

@article{liu2013efficient,
  title={Efficient space-time sampling with pixel-wise coded exposure for high-speed imaging},
  author={Liu, Dengyu and Gu, Jinwei and Hitomi, Yasunobu and Gupta, Mohit and Mitsunaga, Tomoo and Nayar, Shree K},
  journal={IEEE transactions on pattern analysis and machine intelligence},
  volume={36},
  number={2},
  pages={248--260},
  year={2013},
  publisher={IEEE}
}

@article{feng2016per,
  title={Per-pixel coded exposure for high-speed and high-resolution imaging using a digital micromirror device camera},
  author={Feng, Wei and Zhang, Fumin and Qu, Xinghua and Zheng, Shiwei},
  journal={Sensors},
  volume={16},
  number={3},
  pages={331},
  year={2016},
  publisher={MDPI}
}

@inproceedings{vargas2021time,
  title={Time-multiplexed coded aperture imaging: Learned coded aperture and pixel exposures for compressive imaging systems},
  author={Vargas, Edwin and Martel, Julien NP and Wetzstein, Gordon and Arguello, Henry},
  booktitle={Proceedings of the IEEE/CVF International Conference on Computer Vision},
  pages={2692--2702},
  year={2021}
}

@article{zhang2016compact,
  title={Compact all-CMOS spatiotemporal compressive sensing video camera with pixel-wise coded exposure},
  author={Zhang, Jie and Xiong, Tao and Tran, Trac and Chin, Sang and Etienne-Cummings, Ralph},
  journal={Optics express},
  volume={24},
  number={8},
  pages={9013--9024},
  year={2016},
  publisher={Optical Society of America}
}

@article{luo2017exposure,
  title={Exposure-programmable CMOS pixel with selective charge storage and code memory for computational imaging},
  author={Luo, Yi and Ho, Derek and Mirabbasi, Shahriar},
  journal={IEEE Transactions on Circuits and Systems I: Regular Papers},
  volume={65},
  number={5},
  pages={1555--1566},
  year={2017},
  publisher={IEEE}
}

@incollection{raskar2006coded,
  title={Coded exposure photography: motion deblurring using fluttered shutter},
  author={Raskar, Ramesh and Agrawal, Amit and Tumblin, Jack},
  booktitle={Acm Siggraph 2006 Papers},
  pages={795--804},
  year={2006}
}

@article{martel2020neural,
  title={Neural sensors: Learning pixel exposures for HDR imaging and video compressive sensing with programmable sensors},
  author={Martel, Julien NP and Mueller, Lorenz K and Carey, Stephen J and Dudek, Piotr and Wetzstein, Gordon},
  journal={IEEE transactions on pattern analysis and machine intelligence},
  volume={42},
  number={7},
  pages={1642--1653},
  year={2020},
  publisher={IEEE}
}

@article{faraji2019hyperspectral,
  title={Hyperspectral imager with folded metasurface optics},
  author={Faraji-Dana, MohammadSadegh and Arbabi, Ehsan and Kwon, Hyounghan and Kamali, Seyedeh Mahsa and Arbabi, Amir and Bartholomew, John G and Faraon, Andrei},
  journal={Acs Photonics},
  volume={6},
  number={8},
  pages={2161--2167},
  year={2019},
  publisher={ACS Publications}
}

@inproceedings{yu2025active,
  title={Active Hyperspectral Imaging Using an Event Camera},
  author={Yu, Bohan and Liang, Jinxiu and Wang, Zhuofeng and Fan, Bin and Subpa-asa, Art and Shi, Boxin and Sato, Imari},
  booktitle={Proceedings of the Computer Vision and Pattern Recognition Conference},
  pages={929--939},
  year={2025}
}
